\documentclass[12pt]{article}
\pdfoutput=1
\usepackage{epsf}
\usepackage{cite}
\usepackage{color}
\usepackage{graphicx}

\usepackage{amssymb,amsmath}

\newcommand{\ord}{\mathcal{O}}

\newcommand{\IM}{\rm{Im}}
\newcommand{\RE}{\rm{Re}}

\newcommand{\tev}{\, {\rm TeV}}
\newcommand{\gev}{\, {\rm GeV}}
\newcommand{\mev}{\, {\rm MeV}}

\newcommand{\vcb}{|V_{cb}|}
\newcommand{\vtd}{|V_{td}|}
\newcommand{\vub}{|V_{ub}|}
\newcommand{\vts}{|V_{ts}|}
\newcommand{\vus}{|V_{us}|}

\newcommand{\beq}{\begin{equation}}
\newcommand{\eeq}{\end{equation}}
\newcommand{\be}{\begin{equation}}
\newcommand{\ee}{\end{equation}}
\newcommand{\ba}{\begin{array}}
\newcommand{\ea}{\end{array}}
\newcommand{\beqa}{\begin{eqnarray}}
\newcommand{\eeqa}{\end{eqnarray}}
\newcommand{\bea}{\begin{eqnarray}}
\newcommand{\eea}{\end{eqnarray}}
\newcommand{\beqn}{\begin{eqnarray}}
\newcommand{\eeqn}{\end{eqnarray}}

\newcommand{\D}{\Delta}

\newcommand{\eps}{\epsilon}
\newcommand{\nn}{\nonumber}

\newcommand{\lsim}{\stackrel{<}{_\sim}}
\newcommand{\gsim}{\stackrel{>}{_\sim}}

\definecolor{red}{cmyk}{0,1,1,0.4}

\def\kpn{K^+\rightarrow\pi^+\nu\bar\nu}
\def\klpn{K_{L}\rightarrow\pi^0\nu\bar\nu}
\def\epe{\varepsilon'/\varepsilon}

\newcommand{\bsi}{B_6^{(1/2)}}
\newcommand{\bei}{B_8^{(3/2)}}

\begin{document}

\begin{flushright}
 	CERN-PH-TH-2015-158\\
 TTP15-023\\
 {FLAVOUR(267104)-ERC-104}
\end{flushright}

\medskip

\begin{center}
{\Large \bf
Quark flavour observables in the Littlest Higgs \vspace{1mm}\\
model with T-parity after LHC Run 1
}\\[0.8 cm]

{\bf Monika Blanke$^{a,b,c}$, Andrzej~J.~Buras$^{d,e}$ and Stefan Recksiegel$^{e}$} \\[0.5 cm]
{\small
$^a$CERN Theory Division, CH-1211 Geneva 23, Switzerland\vspace{1mm}\\
$^b$Institut f\"ur Theoretische Teilchenphysik, Karlsruhe Institute of Technology,
Engesserstra{\ss}e 7,\\ D-76128 Karlsruhe, Germany\vspace{1mm}\\
$^c$Institut f\"ur Kernphysik, 
Karlsruhe Institute of Technology,
Hermann-von-Helmholtz-Platz 1,\\ D-76344 Eggenstein-Leopoldshafen, Germany\vspace{1mm}\\
$^d$TUM Institute for Advanced Study, Lichtenbergstr. 2a, D-85747 Garching, Germany\vspace{1mm}\\
$^e$Physik Department, Technische Universit\"at M\"unchen,
James-Franck-Stra{\ss}e, \\D-85747 Garching, Germany}
\end{center}

\abstract{%
\noindent
The Littlest Higgs Model with T-parity (LHT) 
belongs to the simplest new physics scenarios with new sources of flavour 
and CP violation. The latter originate in the interactions of ordinary quarks 
and leptons with heavy mirror quarks and leptons that are mediated 
by new  heavy gauge bosons. 
Also a heavy fermionic top partner is present in this model which communicates with 
the SM fermions by means of standard $W^\pm$ and $Z^0$ gauge bosons.
We present a new analysis of quark flavour observables in the LHT model in view of 
the oncoming flavour precision era. We use all available information on the CKM parameters, lattice QCD input and experimental data on quark flavour observables and corresponding theoretical calculations, 
taking into account new lower bounds on the symmetry breaking scale and the mirror quark masses from the LHC. We investigate  by how much the branching ratios for a number of rare $K$ and $B$ decays are still allowed to depart from their SM values. This includes $\kpn$, $\klpn$, $K_L\to \mu^+\mu^-$,  
$B\to X_s\gamma$, $B_{s,d}\to\mu^+\mu^-$, $B\to K^{(*)}\ell^+\ell^-$, $B\to K^{(*)}\nu\bar\nu$, and $\epe$.  Taking into account the constraints from $\Delta F=2$ processes, significant departures from the SM predictions for 
$\kpn$ and $\klpn$ are possible, while the effects in $B$ decays are much smaller.
In particular, the LHT model favours  $\mathcal{B}(B_{s}\to\mu^+\mu^-) \ge  \mathcal{B}(B_{s}\to\mu^+\mu^-)_{\rm SM}$, which is not supported by the data, and the present anomalies in  $B\to K^{(*)}\ell^+\ell^-$ decays cannot be explained in this model. {With the recent lattice and large $N$ input the imposition 
of the $\epe$ constraint implies a significant suppression of the branching 
ratio for $\klpn$ with respect to its SM value while  allowing only for 
small modifications of $\kpn$.}
Finally, we investigate how the LHT physics could 
be distinguished from other models by means of indirect measurements and discuss the consequences for quark flavour observables of not finding any LHT state in the coming years.
}

\newpage

\section{Introduction}
\setcounter{equation}{0}
Elementary Particle Physics stands at the threshold of big discoveries. The completion of the Standard Model (SM) through the Higgs discovery in 2012 \cite{Aad:2012tfa,Chatrchyan:2012ufa} has 
shown that we are on the right track towards the fundamental theory. 
But there is a common belief that in order to understand the nature around us new particles and new forces are required. Fortunately in the coming years the ATLAS and 
CMS experiments will tell us directly  whether new physics (NP) is present 
up to scales as high as several $\tev$. These efforts will be accompanied 
by the indirect search for NP with the  help of quantum fluctuations.
This indirect route to short distance scales will be followed 
in this decade {by several experiments \cite{Hewett:2012ns}, in particular the LHCb experiment and to some extent by CMS and ATLAS 
through more precise data on rare $B_{s,d}$ decays and CP violation. But equally important are
the dedicated kaon experiments NA62 at CERN and KOPIO at J-PARC and the Belle II
experiment at SuperKEKB.}
 Also the study of charged lepton flavour violation
and of electric dipole moments at various laboratories will be very important in this respect.

One of the important questions in this context is whether the 
framework of constrained Minimal Flavour Violation (CMFV) \cite{Buras:2000dm,Buras:2003jf,Blanke:2006ig}
and the more general framework of MFV \cite{D'Ambrosio:2002ex} will be 
capable of describing the future data. In models of this class, 
when flavour blind phases are absent or set to zero, stringent 
relations between various observables in the $K$, $B_d^0$ and $B_s^0$ systems 
 are present \cite{Buras:2003jf}. Consequently 
the  departures from SM expectations in this class of models in these three meson systems 
are correlated with each other, allowing very transparent tests of these 
simple NP scenarios. However, generally these relations can be strongly 
violated, implying often other correlations between observables characteristic 
for a given NP scenario. 
Such correlations, being less 
sensitive to the model parameters than individual observables, can often allow a transparent 
distinction between various models proposed in the literature 
\cite{Buras:2013ooa}. 

Among the simplest extensions of the SM that go beyond the concept 
of MFV is the Littlest Higgs Model with T-parity (LHT) \cite{ArkaniHamed:2002pa,ArkaniHamed:2002qx,ArkaniHamed:2002qy,Cheng:2003ju,Cheng:2004yc}. 
In this model, new heavy fermions and gauge bosons are present. 
The interactions of ordinary quarks 
and leptons with these new heavy mirror quarks and leptons,  mediated 
by new  heavy electroweak gauge bosons, 
introduce new sources of flavour 
and CP violation. The most characteristic signals of these new interactions 
are violations of CMFV and MFV relations between observables in different meson systems.
At the same time, no new effective operators are generated beyond those which are already present in the SM. Therefore non-perturbative uncertainties are not increased with respect to the ones present in the SM. This operator structure can be tested by studying correlations between observables from the same meson system.

In the last decade we have performed a number of extensive phenomenological 
analyses of the LHT model \cite{Blanke:2006sb,Blanke:2006eb,Blanke:2007db,Blanke:2007ee,Blanke:2007wr,Blanke:2008ac,Bigi:2009df,Blanke:2009am}. Further phenomenological discussions of flavour in the LHT model can be found in \cite{Hubisz:2005bd,Goto:2008fj,delAguila:2008zu}. Our 2009 analysis in 
\cite{Blanke:2009am} has shown that significant deviations from SM expectations 
were possible in the LHT model at that time.
Our main findings in 2009, related to quark flavour physics, can be 
summarized as follows:
\begin{itemize}
\item
The CMFV relations between $K$, $B_d$ and $B_s$ 
systems can be strongly violated. This allowed to remove the tension between 
$\varepsilon_K$ and $S_{\psi K_S}$ \cite{Lunghi:2008aa,Buras:2008nn,Buras:2009pj,Lunghi:2009sm,Lunghi:2010gv}.
\item 
 Interestingly, in the LHT model it was not possible to obtain the mixing 
induced CP-asymmetry
$S_{\psi\phi}$ of $\ord(1)$ and  
 values above 0.3 were very unlikely. In fact the most recent data from  
LHCb \cite{Aaij:2014zsa}  confirm this prediction.  Yet the LHT model can both enhance or suppress $S_{\psi\phi}$ w.\,r.\,t.\ its SM value. As we will stress below this could provide an
important distinction from other models, like the Two Higgs Doublet model with MFV and flavour blind phases
(${\rm 2HDM_{\overline{MFV}}}$)  \cite{Buras:2010mh,Buras:2010zm} 
where $S_{\psi\phi}$ can only be enhanced due to its correlation with $S_{\psi K_S}$. 
\item
$\mathcal{B}(\klpn)$ and $\mathcal{B}(\kpn)$ could be enhanced by factors of
 3 and 2.5, respectively, but not simultaneously with $S_{\psi\phi}$. Also, a distinctive correlation between these two branching ratios, typical for models with only SM operators \cite{Blanke:2009pq}, holds.
\item
 Rare $B_{s,d}$ decays turned out to be SM-like but still some measurable departures from SM predictions  
were possible. 
In particular $\mathcal{B}(B_{s,d}\to\mu^+\mu^-)$ could be enhanced 
by $30\%$, with a
significant part of this enhancement coming from the T-even sector.
\end{itemize}

In view of the oncoming flavour precision era it is of interest to 
update our 2009 analysis, as during the last six years substantial improvements on both experimental and theoretical inputs have been achieved. 
%improved data on various flavour observables have been presented and also the input from QCD lattice has been modified. 
In particular:
\begin{itemize}
\item
The data from ATLAS and CMS, both on Higgs physics and on direct NP searches, provide important constraints on the LHT parameter space. Further significant improvements can be expected from LHC Run 2. In particular in our 2009 analysis we had restricted the mirror quark masses to lie below the $1\tev$ scale, in order to make them easily accessible to direct searches. The absence of a signal in run 1 of the LHC however pushes the masses of these fermions to {heavier} ranges \cite{Reuter:2013iya}. As we will see below this change has a significant impact on the possible size of LHT effects in rare decays.
\item
The values of CKM parameters extracted from tree-level decays are 
presently better constrained and will be significantly improved in the 
coming years.
\item
Significant progress has been made by the lattice community in calculating 
various parameters like weak decay constants and  non-perturbative $B_i$  parameters.
\item
The mixing induced CP-asymmetry $S_{\psi\phi}$ is presently  known with much higher accuracy than in 2009. 
\item
The branching ratio $\mathcal{B}(B_{s}\to\mu^+\mu^-)$ has been found 
SM-like, as expected within the LHT model, but significant NP contributions 
are still allowed due to the large experimental uncertainty and to a lesser extent parametric uncertainties  dominantly present in the value  of $\vcb$. Still, the improved precision on the SM prediction for  $\mathcal{B}(B_{s}\to\mu^+\mu^-)$ makes a detailed comparison of theory and data possible.
%\item
% The SM branching ratio for $B\to X_s\gamma$ decay increased, since our analysis in 2006  \cite{Blanke:2006sb}, by $6\%$ towards the experimental value, decreasing significantly the room for NP contributions. Even if the latter contribution were found by to be small in the LHT model, it is interesting to see whether LHT is consistent with the 
% data.
\item
The data on $B\to K^{(*)}\ell^+\ell^-$ from LHCb provided a new arena for testing the LHT model. In fact, it will turn out that the LHT model is unable to 
describe this new data. 
\item
{The measured values of the ratios $R(D)$ and $R(D^*)$ show a $3.9\sigma$ deviation from their SM predictions \cite{Amhis:2014hma}. We will investigate whether the LHT model could be the origin of this discrepancy. Note that these ratios have not been considered in the context of the LHT model before.}
\item
{The new results for the non-perturbative parameters $\bsi$ and $\bei$ from 
lattice QCD \cite{Blum:2015ywa,Bai:2015nea} and the large $N$ approach \cite{Buras:2015xba} imply 
that $\epe$ in the SM is significantly below the data \cite{Buras:2015yba}. The question arises 
whether the LHT model could help in solving this problem.}
\item
Very importantly the NA62 experiment at CERN should provide in the next 
years a new measurement of $\mathcal{B}(\kpn)$ which will be an important 
test of the LHT model in view of very small theoretical uncertainties in this 
decay.
\end{itemize}

In view of these developments  the two main goals of our present analysis are:
\begin{itemize}
\item
We confront the rich pattern of flavour violation in this model with
 the present data and investigate the allowed size of new flavour violating effects, taking  present bounds and improved input into account.
\item
We investigate what size of new flavour violating effects will still be possible if we do not find any LHT state during the next LHC run. This means setting the masses of new gauge bosons and mirror quarks to be several TeV.
\end{itemize}

Our paper is organized as follows. In Section~\ref{sec:model} we recall 
basic features of the LHT model that are relevant to understand our analysis. In particular, we recall the flavour structure of this model. Due to the absence of new operators, the full quark flavour analysis can be formulated in terms of a number of one-loop master functions. We refrain from repeating the complete formulae for these functions in the LHT model that can all be found in our previous papers. But in Section~\ref{sec:basicf} we collect the relevant expressions for quark flavour observables that can be compactly written in terms of these master functions. This will allow us to indicate the changes in the CKM input and in 
non-perturbative parameters as well as QCD corrections that took place since our 2009 analysis.  Section~\ref{sec:bounds} is devoted to a brief review of the direct constraints on the LHT parameter space, implied by the available data from ATLAS and CMS. In Section~\ref{sec:num}, after presenting our strategy 
for the numerical analysis and summarizing the input, we present the results for a multitude of observables in the quark sector. 
The highlights of our analysis 
are listed in Section~\ref{sec:sum}, where we also present a brief outlook 
for the coming years.

\section{General structure of the LHT model}\label{sec:model}

\subsection{Preliminaries}

The Littlest Higgs model
without \cite{ArkaniHamed:2002qy} T-parity 
has been invented to solve the problem of the quadratic 
divergences in the Higgs mass without using supersymmetry.
In this approach the cancellation of divergences in $ m_H$ is achieved with 
the help of new particles of the same spin-statistics. 
Basically the SM Higgs is kept light 
because it is a pseudo-Goldstone boson of a spontaneously broken global 
symmetry:
\be 
SU(5)\to SO(5).
\ee
Thus the Higgs mass is protected by a global symmetry. In order to achieve this the gauge group has to be extended
to
\be
 G_{\rm LHT}=SU(3)_c\times [SU(2)\times U(1)]_1\times[SU(2)\times U(1)]_2
\ee
 and the symmetry breaking mechanism has to be properly arranged
{(\it collective symmetry breaking)}. 
Excellent reviews of Little Higgs models can be found in
\cite{Schmaltz:2005ky,Perelstein:2005ka}.

\subsection{Particle content of the LHT model}

In order to make the Littlest Higgs model
consistent with electroweak precision tests and simultaneously have
the new particles of this model in the reach of the LHC, a discrete symmetry,
T-parity, has been introduced \cite{Cheng:2003ju,Cheng:2004yc}. 
Under T-parity all SM particles are {\it even}.
Among the new particles only a heavy $Q=+2/3$ charged top partner quark, called $T_+$, belongs to the
even sector. Its role is to cancel the quadratic divergence in the Higgs
mass generated by the ordinary top quark. The even sector and also the 
model without T-parity belong to the CMFV class if only flavour violation 
in the down-quark sector is considered \cite{Buras:2004kq,Buras:2006wk}.

More interesting from the point of view of FCNC processes in the quark sector is the T-odd
sector. It contains three doublets of mirror quarks
 \begin{equation}\label{2.6}
\begin{pmatrix} u^1_{H}\\d^1_{H} \end{pmatrix}\,,\qquad
\begin{pmatrix} u^2_{H}\\d^2_{H} \end{pmatrix}\,,\qquad
\begin{pmatrix} u^3_{H}\\d^3_{H} \end{pmatrix}\,.
\end{equation}

To first order in $v/f$, with $f=\ord(1\tev)$, the mirror quarks have
 vectorial
couplings under $SU(2)_L\times U(1)_Y$ and their masses satisfy
\begin{equation}\label{2.7a}
m^u_{H1}=m^d_{H1}\,,\qquad m^u_{H2}=m^d_{H2}\,,\qquad
m^u_{H3}=m^d_{H3}\,.
\end{equation}

Mirror quarks 
communicate with the SM quarks by means of heavy  gauge bosons
\begin{equation}\label{2.3}
W_H^\pm\,,\qquad Z_H\,,\qquad A_H\,,
\end{equation}
which can be considered as ``partners'' of the SM gauge bosons. 
They are T-odd particles 
with masses given to lowest order in $v/f$ by
\begin{equation}\label{2.4}
M_{W_H}=M_{Z_H}=gf\,,\qquad
M_{A_H}=\frac{g'f}{\sqrt{5}}=\frac{\tan{\theta_W}}{\sqrt{5}}M_{W_H}\simeq\frac{M_{W_H}}{4.1}\,,
\end{equation}
where $g$ and $g'$ are the usual couplings of $SU(2)_L$ and $U(1)_Y$, respectively.

\subsection{Flavour structure of the LHT model}

The interactions between ordinary down quarks and mirror quarks, mediated by 
gauge bosons $W_H^\pm\,$, $Z_H\,$, $A_H$, are  
governed by the new mixing matrix $V_{Hd}$.
The corresponding matrix $V_{Hu}$ in the up
sector is obtained by means of the relation
\cite{Hubisz:2005bd,Blanke:2006xr}
\be
V_{Hu}^\dagger V_{Hd}^{\,}=V^{}_\text{CKM}\,.
\ee
 Thus we have new flavour and CP-violating contributions to decay amplitudes
in this model. These new interactions
 can have a structure that is very different from the CKM matrix.

The difference between the CMFV models and the LHT model can be transparently 
seen in the formulation of FCNC processes in terms of the master 
one-loop functions that multiply the CKM factors $\lambda_t^{(i)}$
\be
\lambda_t^{(K)} = V_{ts}^*\,V_{td}\,,\qquad
\lambda_t^{(d)} = V_{tb}^*\,V_{td}\,,\qquad
\lambda_t^{(s)} = V_{tb}^*\,V_{ts}\,,
\label{eq:lambdas}
\ee
for $K$, $B_d$ and $B_s$ systems respectively.
 This formulation can be used straightforwardly here because the LHT model 
has the same operator structure as the SM and the models with CMFV, except that
the real and universal master functions of the latter models
become complex quantities and the property of the flavour universality of these 
functions is lost. Consequently the usual CMFV relations between $K$, $B_d$ and $B_s$ systems are generally  broken. 

Explicitly, the new functions in the 
LHT model are given as 
follows ($i=K,d,s$)
\bea 
S_i &=& S_\text{SM} + \bar S_\text{even} + \frac{1}{\lambda_t^{(i)}}
\bar
S_i^\text{odd} \quad \equiv \quad |S_i|\, e^{i\, \theta_S^i}\,, \label{eq:Si}\\
X_i &=& X_\text{SM} + \bar X_\text{even} + \frac{1}{\lambda_t^{(i)}}
\bar
X_i^\text{odd} \quad\equiv\quad |X_i|\, e^{i\, \theta_X^i}\,,\label{eq:Xi} \\
Y_i &=& Y_\text{SM} + \bar Y_\text{even} + \frac{1}{\lambda_t^{(i)}}
\bar Y_i^\text{odd} \quad\equiv\quad |Y_i|\, e^{i\, \theta_Y^i}\,, \label{eq:Yi}\\
Z_i &=& Z_\text{SM} + \bar Z_\text{even} + \frac{1}{\lambda_t^{(i)}}
\bar Z_i^\text{odd} \quad\equiv\quad |Z_i|\, e^{i\, \theta_Z^i}\,.
\label{eq:Zi} \eea 
Here $S_\text{SM}$, $X_\text{SM}$, $Y_\text{SM}$ and $Z_\text{SM}$ are the SM contributions for which explicit
expressions can be found in \cite{Buras:2013ooa}.  $\bar S_\text{even}$, $\bar
X_\text{even}$, $\bar Y_\text{even}$ and $\bar Z_\text{even}$ are the contributions from the T-even sector, that
is the contributions of $T_+$ and of $t$ at order $v^2/f^2$
necessary to make the  GIM mechanism  work. The latter contributions, similarly to $S_\text{SM}$,
$X_\text{SM}$, $Y_\text{SM}$ and $Z_\text{SM}$, are real and independent of $i=K,
d, s$. Explicit expressions for them can be found in \cite{Blanke:2006sb}.

The
functions  $\bar S_i^\text{odd}$, $\bar X_i^\text{odd}$,  $\bar
Y_i^\text{odd}$ and $\bar Z_i^\text{odd}$  represent the T-odd
sector of the LHT model and are obtained from penguin and box
diagrams with internal mirror quarks and new gauge bosons. 
Explicit expressions for these functions can be found in our previous 
papers \cite{Blanke:2006sb,Blanke:2006eb,Blanke:2009am} and will not be repeated here. 

At this point it should be recalled that in our earlier papers, when calculating  $\bar X_i^\text{odd}$,  $\bar
Y_i^\text{odd}$ and $\bar Z_i^\text{odd}$, we had overlooked an $\mathcal{O}(v^2/f^2)$ contribution to {the} $Z^0$-penguin diagrams.
This contribution has been identified by Goto et al. \cite{Goto:2008fj}
in the context of their study of {the} $K\to\pi\nu\bar\nu$ decays in the LHT model, and independently by del Aguila et al.\ \cite{delAguila:2008zu} in the context  of the corresponding analysis of the LFV decays $\mu\to e\gamma$ and $\mu\to3e$. At the same time, these authors have confirmed our calculations except for the omission mentioned above. The corrected Feynman rules of \cite{Blanke:2006eb} implied by the findings of \cite{Goto:2008fj,delAguila:2008zu} are collected in Appendix A in \cite{Blanke:2009am}. In that paper also the implied shifts
 in the corresponding $Z$-penguin functions and consequently in  $\bar X_i^\text{odd}$,  $\bar Y_i^\text{odd}$ and $\bar Z_i^\text{odd}$ are given.

A review on flavour physics in 
the LHT model can be found in 
\cite{Blanke:2007ww} and selected papers containing details
of the pattern of flavour violation in this model can be found in
\cite{Blanke:2006sb,Blanke:2006eb,Blanke:2007db,Goto:2008fj,delAguila:2008zu,Blanke:2009pq,Blanke:2009am}.

\subsection{LHT as a representative example}

{Before moving on, we addres the question whether our results remain valid in the more general context of Little Higgs models with T-parity, independent of the details of the Littlest Higgs model\footnote{We thank an unknown referee for raising this question.}. The flavour violating effects in the LHT model found by us are mostly due to the T-odd sector of the model, namely the heavy electroweak gauge bosons and mirror fermions, with only left-handed couplings to the SM quarks and leptons. The presence of these states is generic to the class of Little Higgs models with T-parity. Some details, like the precise form of the mirror quark coupling to the standard $Z$ boson, are indeed model dependent, rendering a general quantitative analysis of the whole class of Little Higgs models with T-parity impossible. However we point out that the overall structure of effects remains unaffected. We therefore expect our results to hold, at least qualitatively, beyond the concrete and rather restricted framework of the LHT model.
}

\section{Basic formulae for quark flavour observables}\label{sec:basicf}

\boldmath
\subsection{$\Delta F=2$ Observables}
\unboldmath

The flavour parameters of the quark sector in the LHT model are first of all 
bounded by very precise data on 
\be\label{DF2}
\Delta M_s\,, \qquad \Delta M_d\,, \qquad \varepsilon_K\,,
\ee
but also by the data on the mixing induced CP-asymmetries in $B_d^0\to J/\psi K_S$ and $B_s^0\to J/\psi\phi$ \cite{Amhis:2014hma,Aaij:2014zsa}  \footnote{In our conventions $S_{\psi\phi} = -\sin\phi_s$, with the measured value for $\phi_s$ quoted by LHCb and HFAG.}
\be\label{SDATA}
S_{\psi K_S}= 0.691\pm 0.017,\qquad S_{\psi\phi}= 0.015\pm 0.035 \,.
\ee
Although $S_{\psi\phi}$ is found to be small {it could still significantly differ from its SM value
\be\label{SpsiphiSM}
S_{\psi\phi}^\text{SM}= \sin(2|\beta_s|)=0.036\pm0.002\,.
\ee
The numerical value for
\be\label{SpsiKSSM}
S_{\psi K_S}^\text{SM}=\sin 2\beta
\ee
depends strongly on the value of $|V_{ub}|$, as can be seen from Fig.\ \ref{fig:SpsiKS-KL}.  
Here $\beta$ and  $\beta_s$ are defined by
\be\label{vtd}
V_{td}=\vtd e^{-i\beta}, \qquad V_{ts}=-\vts e^{-i\beta_s}.
\ee
}

In the LHT model the mass differences $\Delta M_s$ and $\Delta M_d$ are 
simply given by
\be\label{DMs}
\Delta M_s =\frac{G_F^2}{6 \pi^2}M_W^2 m_{B_s}\left|\lambda_t^{(s)}\right|^2   F_{B_s}^2\hat B_{B_s} \eta_B |S_s|
\ee
and
\be\label{DMd}
\Delta M_d =\frac{G_F^2}{6 \pi^2}M_W^2 m_{B_d}\left|\lambda_t^{(d)}\right|^2   F_{B_d}^2\hat B_{B_d} \eta_B |S_d|
\ee
with the numerical values of all parameters collected in Table~\ref{tab:input}.

Next, the presence of new sources of CP violation coming from the T-odd sector
modifies
the SM formulae in (\ref{SpsiphiSM}), \eqref{SpsiKSSM} as follows
\begin{equation}
S_{\psi K_S} = \sin(2\beta+2\varphi_{B_d})\,, \qquad
S_{\psi\phi} =  \sin(2|\beta_s|-2\varphi_{B_s})\,.
\label{eq:3.43a}
\end{equation}
Here $\varphi_{B_q}$ are NP phases in $B^0_q-\bar B^0_q$ mixings. They
are  directly given in terms of the phases 
of the loop functions $S_q$:
\be
2\varphi_{B_q}=-\theta_S^q\,.
\ee

The formulae for $\Delta M_K$ and $\varepsilon_K$ are more complicated because 
also charm contributions are present. They can all be found in  \cite{Blanke:2006sb}. The only modification relative to these formulae is the change in the overall multiplicative factor in $\varepsilon_K$
\be
e^{i\pi/4} \rightarrow \kappa_\eps e^{i\varphi_\eps},
\ee
where $\varphi_\eps = (43.51\pm0.05)^\circ$ and $\kappa_\eps=0.94\pm0.02$ \cite{Buras:2008nn,Buras:2010pza} takes into account 
that $\varphi_\eps\ne \tfrac{\pi}{4}$ and includes long distance effects in 
${\rm Im}(\Gamma_{12})$ and ${\rm Im}(M_{12})$.

In what follows
we will present the most interesting branching ratios in terms of the functions 
$X_i$ and $Y_i$. The CKM elements that we will use are those
determined from tree level decays and consequently they are independent of 
new physics.

\subsection{\boldmath${B_{s,d}\to\mu^+\mu^-}$\unboldmath}

Interesting implications on the LHT model arise also from the data on $B_{s,d}\to\mu^+\mu^-$. The most recent prediction in the SM  that includes NNLO QCD corrections  \cite{Hermann:2013kca}  and 
NLO electroweak corrections  \cite{Bobeth:2013tba}, put together in \cite{Bobeth:2013uxa}, and the 
most recent averages from the combined analysis of CMS and LHCb \cite{CMS:2014xfa} are given as follows: 
\begin{align}\label{LHCb2}
&\overline{\mathcal{B}}(B_{s}\to\mu^+\mu^-)_{\rm SM} = (3.65\pm0.23)\cdot 10^{-9},\quad
&&\overline{\mathcal{B}}(B_{s}\to\mu^+\mu^-)_\text{exp} = (2.8^{+0.7}_{-0.6}) \cdot 10^{-9}, 
\\
\label{LHCb3}
&\mathcal{B}(B_{d}\to\mu^+\mu^-)_{\rm SM}=(1.06\pm0.09)\cdot 10^{-10}, \quad
&&\mathcal{B}(B_{d}\to\mu^+\mu^-)_\text{exp} =(3.9^{+1.6}_{-1.4})\cdot 10^{-10}. \quad
\end{align}
The ``bar'' in the case of $B_{s}\to\mu^+\mu^-$ indicates the flavour averaged branching ratio, i.\,e.\ $\Delta\Gamma_s$ 
effects \cite{DescotesGenon:2011pb,DeBruyn:2012wj,DeBruyn:2012wk} have been taken into account in the SM prediction.

As we will be using CKM elements determined in tree-level decays, it is 
useful to consider the ratios
\bea
 \mathcal R_s^{\mu\mu}= \frac{\overline{\mathcal{B}}(B_s\to\mu^+\mu^-)}{\overline{\mathcal{B}}(B_s\to\mu^+\mu^-)_\text{SM}}&=&
\left|\frac{Y_s}{Y_\text{SM}}\right|^2 r(\Delta\Gamma_s),\label{eq:Bsmumu} \\
 \mathcal R_d^{\mu\mu}= \frac{\mathcal{B}(B_d\to\mu^+\mu^-)}{\mathcal{B}(B_d\to\mu^+\mu^-)_\text{SM}}&=&
\left|\frac{Y_d}{Y_\text{SM}}\right|^2,\label{eq:Bdmumu}
\eea
so that the leading dependence on CKM factors cancels out in these ratios. However, a residual CKM dependence is present in the shifts due to contributions from the T-odd sector, as seen  in (\ref{eq:Yi}). The factor $r(\Delta\Gamma_s)$ represents the difference between $\Delta\Gamma_s$ effects in the LHT model and in the SM. 
Using the general formulae in \cite{Buras:2013uqa} we find in the LHT model
\be
r(\Delta\Gamma_s)=\frac{1+y_s\cos(2\theta^s_Y-2\varphi_{B_s})}{1+y_s}\,,
\ee
where \cite{Amhis:2014hma}
\be
y_s=\frac{\Delta\Gamma_s}{2\Gamma_s}=0.061\pm 0.005 \,.
\ee
 We find that in the LHT model $r(\Delta\Gamma_s)$ deviates from unity by at most 0.5\% and can therefore be set to unity.

 The ratios $\mathcal R_{s,d}^{\mu\mu}$
are independent of the meson weak decay constants. The relevant SM expressions for these branching 
ratios can be found in  \cite{Buras:2013uqa}. Using these expressions together 
with (\ref{eq:Bsmumu}) and  (\ref{eq:Bdmumu}) the corresponding results for the 
LHT model can be found.

While the ratios in question show transparently the size of departures from 
the SM predictions independently of the values of weak decay constants and 
CKM parameters, they hide these parametric uncertainties present both in the 
SM and the LHT model. In particular, both branching ratios depend quadratically 
on the value of $\vcb$. The authors in \cite{Bobeth:2013uxa} used the inclusive value for $\vcb\approx 42.2\times 10^{-3}$ and obtained the SM result $B_s\to\mu^+\mu^-$  in (\ref{LHCb2}) that is 
by  $1.2\sigma$  above the data. 
For the exclusive determinations of $\vcb$, as known presently, the SM would 
be much closer  to the data.
%But, the very recent Belle data, moved the exclusive value upwards and
% the SM value in (\ref{LHCb2}) could eventually turn out to 
%represent well the SM prediction for $B_s\to\mu^+\mu^-$.}

From the point of view of the LHT model it is rather crucial to find out whether the SM prediction is indeed higher than the data or not. Indeed, as we will find in Section~\ref{sec:num}, the LHT model favours a slight enhancement of 
$\overline{\mathcal{B}}(B_{s}\to\mu^+\mu^-)$ over its SM value, while 
the data, as seen in (\ref{LHCb2}), favours a moderate suppression. Only a further  improvement on the value of $\vcb$ and the relevant weak decay constants and most 
importantly future more accurate data can tell us whether indeed this 
is a true problem for the LHT model.

\subsection{\boldmath${B\to X_s\gamma}$\unboldmath}

{The most recent NNLO estimate in the SM gives \cite{Czakon:2015exa}
\be
\mathcal{B}(B\to X_s\gamma)_\text{SM} =(3.36\pm0.23)\cdot 10^{-4}\,,
\ee
which agrees very well with the most recent experimental  world average
\be
\mathcal{B}(B\to X_s\gamma)_\text{exp} =(3.43\pm0.22)\cdot 10^{-4}\,.
\ee

The branching ratio for $B\to X_s\gamma$ decay in the LHT model can 
be found in \cite{Blanke:2006sb}.  NP effects in this decay turned out 
to be at the few percent level. Therefore although the room for NP contributions to this decay decreased since 2006, the $B\to X_s\gamma$ branching ratio still does not pose a relevant constraint, beyond those from $\Delta F=2$ observables, on the LHT parameter space.}
On the other hand  
the fact that in this particular case NP effects have been predicted already in 
2006 to be small could be regarded as a success of the LHT model. It remains to be seen whether the 
improvements in the theoretical and experimental accuracy of theory and experiment in this decade will change 
this picture.

\subsection{\boldmath${B\to K^{(*)}\nu\bar\nu}$\unboldmath}

Of interest are also the exclusive $b\to s \nu\bar\nu$ transitions that are theoretically
rather clean and should be measured by Belle II at the end of this decade.
The most recent SM estimates of the relevant branching ratios  \cite{Buras:2014fpa} read:
\be
\mathcal{B}(B^+\to K^+\nu\bar\nu)_\text{SM} = \left[\frac{\vcb}{0.0409}\right]^2(4.0 \pm 0.4) \cdot 10^{-6}, 
\ee
\be
\mathcal{B}(B^0\to K^{* 0}\nu\bar\nu)_\text{SM}  = \left[\frac{\vcb}{0.0409}\right]^2 (9.2\pm 0.9) \cdot 10^{-6},
\ee
where the errors in the parentheses are fully dominated by form factor uncertainties. We expect that when these two branching ratios will be measured, these 
uncertainties will be further decreased and $\vcb$ will be precisely known so 
that a very good test of the SM will be possible.

Again  the ratios between the LHT and SM predictions for these 
 branching ratios are very simple
\bea
 \mathcal R^{\nu\nu}_{K} &=& \frac{\mathcal{B}(B\to K\nu\bar\nu)}{\mathcal{B}(B\to K\nu\bar\nu)_\text{SM}}=
\left|\frac{X_s}{X_\text{SM}}\right|^2,\\
       \mathcal R^{\nu\nu}_{K^*} &=& \frac{\mathcal{B}(B\to K^*\nu\bar\nu)}{\mathcal{B}(B\to K^*\nu\bar\nu)_\text{SM}}                                            =
\left|\frac{X_s}{X_\text{SM}}\right|^2,
\eea
Note that similar to models with CMFV these two ratios are equal to each other, 
which constitutes an important test of the  LHT model. This is related to the 
absence of right-handed {flavour changing} currents in this model.

\subsection{\boldmath $R(D)$ and $R(D^*)$}

{The ratios $R(D)$ and $R(D^*)$, defined as
\be
R(D)=\frac{\Gamma(B\to D\tau\nu)}{\Gamma(B\to D\ell\nu)}\,,\qquad
R(D^*)=\frac{\Gamma(B\to D^*\tau\nu)}{\Gamma(B\to D^*\ell\nu)}\,,
\ee
test the lepton flavour universality in charged current interactions. 
The recent HFAG average \cite{Amhis:2014hma} of BaBar \cite{Lees:2013uzd}, Belle \cite{Huschle:2015rga} and LHCb \cite{Aaij:2015yra} data
\be
R(D)_\text{exp}=0.391\pm 0.041 \pm 0.028 \,,\qquad R(D^*)_\text{exp} = 0.322 \pm 0.018 \pm 0.012
\ee
shows a $3.9\sigma$ deviation from the SM prediction \cite{Kamenik:2008tj,Fajfer:2012vx}
\be
R(D)_\text{SM}=0.297 \pm 0.017\,,\qquad R(D^*)_\text{SM} = 0.252 \pm 0.003\,.
\ee
It is interesting to note that the enhancement with respect to the SM values appears to be universal in both ratios.

Taking a look at the particle content of the LHT model, one might naively hope that this model is able to resolve the anomaly.
It has been shown in a model independent way that a possible solution is the presence of a left-handed charged current contribution \cite{Tanaka:2012nw,Greljo:2015mma}, mediated by a heavy $W'$ boson. For $f\sim 1\tev$ the new gauge boson $W_H$ is in the right mass range. However due to T-parity, the new LHT gauge bosons do not couple to SM fermion pairs. Consequently there are no new tree level contributions to charged current interactions in this model. A new contribution to $R(D)$ and $R(D^*)$ can arise at the one loop level, however the loop suppression together with the smallness of lepton flavour universality breaking effects make it much too small to explain the current $R(D)$ and $R(D^*)$ anomaly.
}

\subsection{\boldmath${K\to\pi\nu\bar\nu}$\unboldmath}

The branching ratios for $\kpn$  and $\klpn$ in the LHT model are given as 
follows
\begin{equation}\label{bkpnn}
\mathcal{B}(K^+\to\pi^+\nu\bar\nu)=\kappa_+ \cdot
\left[\left(\frac{{\rm Im}X_{\rm eff}}{\lambda^5}\right)^2+
\left(\frac{{\rm Re}\lambda_c}{\lambda}P_c(X)+
\frac{{\rm Re}X_{\rm eff}}{\lambda^5}\right)^2\right]\,,
\end{equation}
\begin{equation}\label{bklpn}
\mathcal{B}(\klpn)=\kappa_L\cdot
\left(\frac{{\rm Im}X_{\rm eff}}{\lambda^5}\right)^2\,,
\end{equation}
where  \cite{Mescia:2007kn}
\begin{equation}\label{kapp}
\kappa_+={ (5.173\pm 0.025 )\cdot 10^{-11}\left[\frac{\lambda}{0.225}\right]^8} \,,
\end{equation}
\begin{equation}\label{kapl}
\kappa_L=
(2.231\pm 0.013)\cdot 10^{-10}\left[\frac{\lambda}{0.225}\right]^8
\end{equation}
and $\lambda = |V_{us}|$.  For the charm contribution, represented by $P_c(X)$, the calculations in 
\cite{Buras:2005gr,Buras:2006gb,Brod:2008ss,Isidori:2005xm,Mescia:2007kn}
imply \cite{Buras:2015qea}
\be\label{PCFINAL}
P_c(X)= 0.404\pm 0.024,
\ee
where the error is dominated by the  long distance uncertainty estimated in \cite{Isidori:2005xm}. In what follows we will assume that NP does not modify this value, which turns 
out to be true in all known to us extensions of the SM including the LHT model. Such contributions can be in any case  absorbed into the function $X_{\rm eff}$. The latter function 
that describes pure  short distance contributions from top quark exchanges and 
NP contributions in the LHT model is given by 
\be\label{XK}
X_{\rm eff} = V_{ts}^* V_{td} X_K\,.
\ee

The most recent SM predictions for the branching ratios read \cite{Buras:2015qea}
\begin{align}
\mathcal{B}(\kpn)_\text{SM} &= \left(9.11\pm 0.72\right) \cdot 10^{-11},\label{kpnnfinal}\\
\mathcal{B}(\klpn)_\text{SM} &= \left(3.00 \pm 0.31\right) \cdot 10^{-11}\label{klfinal}\,.
\end{align}

 Experimentally we have \cite{Artamonov:2008qb}
\be\label{EXP1}
\mathcal{B}(\kpn)_\text{exp}=(17.3^{+11.5}_{-10.5})\cdot 10^{-11}\,,
\ee
and the $90\%$ C.L. upper bound \cite{Ahn:2009gb}
\be\label{EXP2}
\mathcal{B}(\klpn)_\text{exp}\le 2.6\cdot 10^{-8}\,.
\ee
Important improvements on these values are expected from the NA62 experiment at CERN
in 2018
\cite{Rinella:2014wfa,Romano:2014xda}, and from the 
measurement of $\klpn$  by KOTO around 2020 at J-PARC \cite{Komatsubara:2012pn,Shiomi:2014sfa}.

\boldmath
\subsection{$K_L\to \mu^+\mu^-$ }\label{sec:5}
\unboldmath

This decay often constrains the size of NP contributions to $\kpn$.
Only the so-called short distance (SD)
part to a dispersive contribution
to $K_L\to\mu^+\mu^-$ can be reliably calculated. It is given 
generally as follows ($\lambda=|V_{us}|=0.2252$):
\be
\mathcal{B}(K_L\to\mu^+\mu^-)_{\rm SD} =
 2.01\cdot 10^{-9} \,
\left ( \frac{{\rm Re} Y^K_{\rm eff} }{\lambda^5} 
  - \bar P_c(Y)  \right )^2  \,,
\ee
where at NNLO \cite{Gorbahn:2006bm}
\be
\bar P_c\left(Y\right) \equiv \left(1-\frac{\lambda^2}{2}\right)P_c\left(Y\right)\,,\qquad
P_c\left(Y\right)=0.115\pm 0.017~.
\ee

The SD contributions in the LHT model are described by
\be\label{YK}
Y^K_{\rm eff} = V_{ts}^* V_{td} Y_{K}
\ee
with
\be
Y_{\rm SM} = \eta_Y Y_0(x_t), \qquad \eta_Y=0.9982
\ee
also entering the $B_{s,d}\to\mu^+\mu^-$ decays. $Y_0(x_t)$ can be found in \cite{Buras:2013ooa} and $\eta_Y$ summarizes both QCD and electroweak corrections 
 \cite{Bobeth:2013uxa}.
%, but to use this value $m_t(m_t)$ has to be inserted 
%into $Y_0(x_t)$.

{As the long-distance contributions to $K_L\to \mu^+\mu^-$ are under poor theoretical control, only a conservative upper bound
\be
\mathcal{B}(K_L\to\mu^+\mu^-)_{\rm SD} < 2.5 \cdot 10^{-9}
\ee
can be derived \cite{Isidori:2003ts}.}

\boldmath
\subsection{$\epe$}\label{sec:epe}
\unboldmath

\subsubsection{SM Contribution}

The starting point of our presentation is the  analytic formula for $\epe$ within the SM \cite{Buras:2003zz,Buras:2015yba}
\be 
\RE(\epe)_{\rm SM}= \IM\lambda_t
\cdot F^\text{SM}_{\varepsilon'}
\label{epeth0}
\ee
with  
\be
F_{\varepsilon'}^\text{SM} =P_0 + P_X \, X_\text{SM} + 
P_Y \, Y_\text{SM} + P_Z \, Z_\text{SM}+ P_E \, E_\text{SM}~.
\label{FE0}
\ee
The first term in (\ref{FE0}) is dominated by QCD-penguin contributions, the next three 
terms by electroweak penguin contributions and the last term is
totally negligible. 

Complete information relevant for our analysis can  be found in 
Appendix B of {\cite{Buras:2015yba}}.
In particular, the coefficients $P_i$ are given in terms of the non-perturbative parameters
\be\label{RS}
R_6\equiv \,\bsi\left[ \frac{114.54\mev}{m_s(m_c)+m_d(m_c)} \right]^2,
\qquad
R_8\equiv \,\bei\left[ \frac{114.54\mev}{m_s(m_c)+m_d(m_c)} \right]^2.
\ee
 as follows:
\begin{equation}
P_i = r_i^{(0)} + 
r_i^{(6)} R_6 + r_i^{(8)} R_8 \,.
\label{eq:pbePi}
\end{equation}
The coefficients $r_i^{(0)}$, $r_i^{(6)}$ and $r_i^{(8)}$ comprise
information on the Wilson-coefficient functions of the $\Delta S=1$ weak
effective Hamiltonian at the NLO. Their numerical values  for 
three  values of $\alpha_s(M_Z)$ are collected 
in Appendix~B of {\cite{Buras:2015yba}}.

In our numerical analysis 
we will use for the quark masses the values 
\cite{Aoki:2013ldr}
\be
m_s(2\gev)=(93.8\pm2.4) \mev, \qquad
m_d(2\gev)=(4.68\pm0.16)\mev.
\ee
Then at the nominal value $\mu=m_c=1.3\gev$ used in {\cite{Buras:2015yba}}, we have 
\be
m_s(m_c)=(109.1\pm2.8) \mev, \qquad
m_d(m_c)=(5.44\pm 0.19)\mev.
\ee

Concerning the parameters $\bsi$ and $\bei$ 
significant progress has been made since our 2007 analysis \cite{Blanke:2007wr}. The
RBC-UKQCD collaboration  \cite{Blum:2015ywa} determined rather precisely the 
value of $\bei$, which transformed to the NDR scheme and the scale $\mu=m_c$, 
reads \cite{Buras:2015qea}
\be\label{B8LATTICE}
B_8^{(3/2)}(m_c)=0.76\pm 0.05\, \qquad \text{(RBC-UKQCD)}
\ee

There is no precise result on $\bsi$ from lattice QCD. From the most recent 
 results of the RBC-UKQCD collaboration \cite{Bai:2015nea} the value of $\bsi$ has 
recently been  extracted {\cite{Buras:2015yba,Buras:2015xba}}
 \be\label{B6LATTICE}
B_6^{(1/2)}(m_c)=0.57\pm 0.19\, \qquad \text{(RBC-UKQCD)}\,.
\ee

But also progress has been made in the large $N$ approach of \cite{Buras:2014maa} (dual QCD) in which in the large $N$ limit one has $\bsi=\bei=1.$ As the recent analysis 
shows one can derive the bounds {\cite{Buras:2015xba}}
\be\label{BG15}
\bsi\le\bei< 1.0\,.
\ee
Moreover, while $\bei$ is found in the ballpark of $0.80\pm 0.10$, 
$\bsi$ is generally smaller and close to the lattice result in (\ref{B6LATTICE}) but the uncertainties are rather large. 

Probably the most important finding of {\cite{Buras:2015xba}} is the bound in (\ref{BG15})
which implies an upper bound on $\epe$ in the SM. Moreover, it has been shown 
that the pattern of the size of various matrix elements in this approach is
supported by the lattice results in  \cite{Bai:2015nea}.

In a very recent paper {\cite{Buras:2015yba}} a new analysis of $\epe$ in the SM has been 
performed assuming that the ${\rm Re}A_0$ and ${\rm Re}A_2$ amplitudes are dominated
by the SM dynamics. In this manner one could determine the matrix elements 
of QCD and electroweak penguin $(V-A)\otimes (V-A)$ operators from the 
precise data on  ${\rm Re}A_0$ and ${\rm Re}A_2$ with much higher precision 
than it is possible presently from lattice QCD. The outcome of this analysis 
is the formula for $\epe$ in (\ref{epeth0}) which is given in terms of $\bsi$ and $\bei$. 

{Using the upper bound in (\ref{BG15}), 
$\bsi\le \bei < 1.0$, one finds,  varying all other parameters within their $1\sigma$ ranges 
{\cite{Buras:2015yba}},
\be\label{SMbound}
\RE(\epe)_{\rm SM}\le \left[\frac{\IM\lambda_t}{1.4\times 10^{-4}}\right]\, (8.6\pm 3.2) \cdot 10^{-4} \,,
\ee
roughly by $2\sigma$  below the experimental result
 \cite{Agashe:2014kda,Batley:2002gn,AlaviHarati:2002ye,Worcester:2009qt}
\be
\RE(\epe)_\text{exp}=(16.6\pm 2.3)\cdot 10^{-4}.
\ee
Using instead the input from lattice QCD the values for $\epe$ in the SM are 
much lower \cite{Buras:2015yba}.}
We will investigate in Section~\ref{sec:num}, whether the LHT model could 
help to remove this discrepancy between the theory and data.

\subsubsection{LHT}

The formula for $\epe$ in the LHT model reads  \cite{Blanke:2007wr}
\be 
{\RE(\epe)} = |\lambda_t|\, \tilde F_{\varepsilon'},
\label{epeth}
\ee
with
\bea
\tilde F_{\varepsilon'} &=&P_0 \sin(\beta-\beta_s) + P_E \,
|E_K|\sin\beta_E^K\nn\\
&& + P_X \, |X_K|\sin\beta_X^K + 
P_Y \, |Y_K|\sin\beta_Y^K
 + P_Z \, |Z_K|\sin\beta_Z^K\,,
\label{FE}
\eea
where 
\be
\beta_i^K = \beta -\beta_s - \theta^K_i\qquad (i = X,Y,Z,E)\,.
\ee
{The coefficients} $P_i$ are the same as in the SM.

\boldmath
\subsection{LHT model facing anomalies in $b\to s \ell^+\ell^-$ transitions}\label{sec:bsll}
\unboldmath

The recent highlights in quark flavour physics were the departures of 
the data on $B_d\to K^{(*)}\mu^+\mu^-$ from the SM expectations, and it is of interest to see how the LHT model faces this data. To this end we recall the shifts caused by NP contributions in the Wilson 
coefficients $C_9$ and $C_{10}$ of the operators 
\be
Q_9 = (\bar s\gamma_\mu P_L b)(\bar \ell\gamma^\mu\ell),\qquad
Q_{10}  = (\bar s\gamma_\mu P_L b)(\bar \ell\gamma^\mu\gamma_5\ell)
\ee
in the LHT model. They are
\begin{align}\label{C9LHT}
 \sin^2\theta_W C_9^\text{NP} &=\Delta Y_s-4\sin^2\theta_W \Delta Z_s,\\
   \sin^2\theta_W C^\text{NP}_{10} &= -\Delta Y_s \label{Yeff}\,.
 \end{align}
Here,
\be 
 \Delta Y_s=Y_s-Y_\text{SM},\qquad \Delta Z_s= Z_s-Z_\text{SM}.
\ee
They can be found by using formulae (\ref{eq:Yi}) and  (\ref{eq:Zi}).

The present anomalies in the angular observables in $B_d\to K^*\mu^+\mu^-$ 
 and the suppression of the branching ratio for $B_d\to K\mu^+\mu^-$ below 
the SM prediction as well as the data on $B_s\to\mu^+\mu^-$ can be well described by \cite{Descotes-Genon:2014uoa,Altmannshofer:2014rta,Hiller:2014yaa,Altmannshofer:2015sma}
\be\label{910}
C_9^\text{NP}\approx -C^\text{NP}_{10} \approx -(0.5\pm 0.2) \,.
\ee
The solution with NP being 
present only in $C_9$ is even favoured, but much harder to explain in the context of  existing models. We refer to \cite{Altmannshofer:2015sma} for tables 
with various solutions and a collection of references to recent papers. 

While the anomalies in $B_d\to K^*\mu^+\mu^-$  are subject to theoretical uncertainties, much cleaner is the 
ratio
\be
 \mathcal R^{\mu e}_K= \frac{\mathcal{B}(B^+\to K^+\mu^+\mu^-)^{[1,6]}}{\mathcal{B}(B^+\to K^+ e^+ e^-)^{[1,6]}}=0.745^{+0.090}_{-0.074}(\text{stat})\pm 0.036(\text{syst})
\label{RLF}\,,
\ee
where the quoted value is the one from LHCb \cite{Aaij:2014ora}. It is by $2.6\sigma$ lower than its SM value $1 +\ord(10^{-4})$ and is an intriguing signal of the breakdown of lepton flavour universality.

All these anomalies turn out to be a problem for the 
LHT model. The relation (\ref{910}) is badly  violated in the LHT model, 
where due to the smallness of the muon vector coupling in the $Z$ penguin  
$C_9^\text{NP}$ turns out to be by an order of magnitude smaller than 
$C_{10}^\text{NP}$. Moreover, $C_{10}^\text{NP}<0$, in variance with (\ref{910}), is favoured in the LHT model. This is the origin of the enhancement of  $B_s\to \mu^+\mu^-$ in this model mentioned above.  
In addition the breakdown of lepton universality in the LHT model is 
absent at the tree-level and even if it can be  generated 
at one loop level, it is by far too small to explain the result in (\ref{RLF}).

Thus, {these anomalies, if confirmed by future more accurate data, have the power to exclude  
 the LHT model as the source of the observed 
pattern of departures from SM expectations for  $b\to s \ell^+\ell^-$ transitions.}

\subsection{\boldmath $D^0-\bar D^0$ mixing}

LHT contributions to $D^0-\bar D^0$ mixing and CP violation have been investigated in detail in \cite{Blanke:2007ee,Bigi:2009df}. In the present paper we refrain from repeating this analysis, however we would like to briefly comment on how the situation changed since 2009.

As in 2009, $D^0-\bar D^0$ mixing in the SM is still plagued by significant hadronic uncertainties. The latter prevent us from obtaining clean correlations between $K$ and $D$ meson observables in the LHT model, which, a priori, are expected in models with only left-handed currents \cite{Blum:2009sk}. The improved experimental constraints on CP violation in $D^0-\bar D^0$ mixing \cite{Amhis:2014hma} therefore do not have a relevant impact on our results for $K$ and $B_{d,s}$ physics observables, which we also confirmed numerically.

\section{Constraints on the LHT parameter space}\label{sec:bounds}

The previous two sections summarized the expressions for flavour observables to be used in our numerical analysis. But, in addition, experiments from the various areas of particle physics place strong constraints on the parameter space of the LHT model and they have to be taken into account.
While the indirect constraints from electroweak precision (EWP) physics are largely unchanged with respect to our earlier analyses, major improvements have been achieved on direct bounds thanks to the first LHC run. Additionally, the discovery of the Higgs boson and the measurement of {its mass as well as} its production and decay rates yields new and partly complementary input. A major analysis of current constraints on the LHT parameter space has been presented in \cite{Reuter:2013iya}. In what follows we briefly recapitulate the new LHT parameters relevant for our analysis and review the current constraints.

\subsection{Electroweak and top sector}

In the electroweak sector the only new parameter is the scale $f$ at which the $SU(5)\to SO(5)$ global symmetry breaking takes place. It determines the mass of the new heavy gauge bosons and scalars and sets the mass scale for the new fermions. 

In the top sector the parameter $x_L$ describes the mixing between the top quark and its T-even partner $T_+$. It also determines the masses of the $T_+$ and $T_-$ quarks, the latter of which is not relevant for FCNC processes.
These parameters are most stringently constrained indirectly, namely from EWP and Higgs data.

EWP constraints on the LHT model have been studied in detail in \cite{Hubisz:2005tx}, and in the context of a simplified model in \cite{Berger:2012ec}. {Recently these analyses have been updated in \cite{Reuter:2013iya}, including the measured value of the Higgs mass $m_h \sim 125\gev$ as well as the T-odd fermion contributions. Interestingly the performed $\chi^2$ fit  showed that scales as low as $\sim 400\gev$ are still consistent with EWP data if the parameter $x_L$, describing the mixing between the top quark and its partner $T_+$, is close to 0.5.}

The bound on the symmetry breaking scale $f$ however increases significantly when the LHC Higgs data are taken into account. Higgs searches alone constrain the scale $f$ to be above $\sim 600\gev$, independently of the parameter $x_L$ \cite{Reuter:2013iya}.

Combining electroweak and Higgs physics constraints yields the lower bound  \cite{Reuter:2013iya}
\be
f\gsim 694\gev\qquad \text{at 95\% C.L.}
\ee
with $x_L\simeq 0.5$. This corresponds to a fine-tuning of at least 5\%.

Interestingly the choice
\be
f = 1\tev\,,\qquad
x_L = 0.5
\ee
we had made in our earlier analyses \cite{Blanke:2006sb,Blanke:2006eb,Blanke:2009am} is still consistent with the currently available indirect constraints. Note that this choice fixes
\be
m_{T_+} = 1.4\tev
\ee
which is still well beyond direct limits from the LHC.

\subsection{Mirror quark sector}

The majority of new parameters in the LHT model is intimately tied to the flavour sector. They arise from the mass matrices of mirror quarks and leptons. Only 
the mass matrix for mirror quarks is relevant in the present paper. It introduces  nine new parameters that can be conveniently divided into the three masses\footnote{Note that the mirror fermions {in a doublet} are degenerate in mass, up to a small splitting from electroweak symmetry breaking.} and a flavour mixing matrix $V_{Hd}$ with three angles and three CP violating phases.
These are
\begin{equation}\label{2.16}
m^q_{H1}\,,\quad m^q_{H2}\,,\quad m^q_{H3}\,,\qquad \theta_{12}^d\,,\quad \theta_{13}^d\,,\quad \theta_{23}^d\,,\qquad \delta_{12}^d\,,\quad \delta_{13}^d\,,\quad \delta_{23}^d\,,
\end{equation}
where the last six parametrise the matrix $V_{Hd}$ in terms of the parametrisation presented in \cite{Blanke:2006xr}. 

\subsubsection{Bounds on mirror quark masses}

The most stringent bounds on the LHT mass spectrum from the LHC experiments are on the mirror quarks, due to their strong coupling to quarks and gluons. Similarly to squarks in supersymmetry, they are pair produced by strong interactions and lead to missing energy signatures with jets and possibly leptons in the final state.

In an early analysis \cite{Perelstein:2011ds} the CMS search for jets and missing transverse energy was used to derive the expected bound
$m^q_{H} \gsim 650\gev$ for $1\,\text{fb}^{-1}$ of data at $\sqrt{s}=7\tev$. 
By now a significantly higher integrated luminosity is available, and many squark searches with different final states have been presented by ATLAS and CMS. The searches most sensitive to LHT mirror quarks have been recasted in \cite{Reuter:2013iya}. Interestingly the most stringent constraints have been found to arise from the search for jets, leptons and missing energy, since the mirror quarks dominantly decay into the heavy gauge bosons $W_H^\pm$, $Z_H$ subsequently producing final state leptons. Assuming a degenerate mirror quark spectrum, for $f=1\tev$ the lower bound
\be\label{eq:mH-bound}
m^q_{H}\gsim 1600\gev
\ee
has been obtained. 

It should be stressed that the bounds on individual mirror quarks can be {weaker} if the requirement of degeneracy is lifted, similarly to the case of non-degenerate squarks \cite{Mahbubani:2012qq}. Furthermore the presence of flavour mixing between the various generations affects the constraints \cite{Blanke:2013uia}.

Upper bounds on the mirror fermion masses can be obtained from their non-decoupling contribution to four-fermion operators \cite{Hubisz:2005tx}.
The constraint on the mirror fermion masses scales linearly with the scale $f$. For $f=1\tev$, and assuming degenerate mirror fermions, the current bound from LEP and LHC data \cite{Reuter:2013iya} is roughly
\be
 m_{H}\lsim 4.6\tev\,.
\ee

\subsubsection{Constraints on mixing parameters}

In contrast to the other LHT parameters, the parameters of the mixing matrix $V_{Hd}$ cannot be constrained by determining the mass spectrum of new particles. However they will in principle be accessible to the LHC by measuring the decays of the mirror quarks into the various SM flavours. Such measurements of branching ratios and CP asymmetries would indeed allow for the most direct determination of mixing angles and CP violating phases in the mirror sector. Similarly to the determination of the CKM matrix from tree level decays, such a method gives the most direct access to the parameters in question. 

This task will however be challenging if not impossible to accomplish at the LHC. Luckily FCNC processes come to the rescue here. 
Even with their help the determination of all these flavour mixing parameters is clearly a very difficult task, in 
particular if no LHT particles will be discovered at the LHC. On
the other hand if in the second round of LHC operation  new
particles present in the LHT model will be  discovered, we
will be able to determine $f$ from $M_{W_H}$, $M_{Z_H}$ or $M_{A_H}$ and $x_L$
from $m_{T_-}$ or $m_{T_+}$. Similarly the mirror fermion masses $m_{Hi}$ will be measured.

Since the CKM parameters can be determined independently of the
LHT contributions from tree level decays during the flavour precision era,
the only remaining free parameters in the quark sector are $\theta_{ij}^d$ and $\delta_{ij}^d$. They can, 
similarly to the parameters of the CKM matrix, {be determined with the help of loop induced 
flavour violating processes. How this determination of the matrix $V_{Hd}$   from loop induced decays would be realised  in practice has  already been discussed in \cite{Blanke:2006sb,Blanke:2006eb} and we will not repeat it here.}

\subsection{Parameter choices for our analysis}\label{Parameters}

In our analysis we will study two different scenarios for the LHT mass scales. 

\paragraph{Scenario A} 
The first one assumes a low new physics scale 
\be
f=1\tev\,,
\ee 
in the reach of the LHC. A low value is clearly preferred by naturalness arguments. In order to optimise the agreement with EWP data, as in our earlier analyses we set the mixing parameter 
\be
x_L = 0.5\,.
\ee 
The mirror quark masses will be varied in the range $(i=1,2,3)$
\be
1600\gev < m^q_{Hi} < 4500\gev
\ee 
in agreement with the current constraints. 

In this context we recall that the T-odd contributions to FCNC processes are governed by the exchange of
mirror fermions and the new gauge bosons in loop diagrams. Consequently the mass splittings between mirror fermions belonging to different doublets are strongly bounded by FCNC processes in correlation with the departure of the matrix $V_{Hd}$  from the unit matrix.

\paragraph{Scenario B}\label{sec:sceB}

The second scenario studies the pessimistic case that no new particles will be found at the LHC in the coming years and no clear deviations from the SM predictions for 
EWP observables will be found. 
Our goal here will then be to find out how large deviations from SM predictions will still be allowed, with the hope that  some deviations from SM predictions in FCNC observables will be detected.
Lacking a detailed analysis of the LHC reach, clearly we can only guess what the bounds on the LHT scales will then be.

The improved knowledge of EWP and Higgs observables will push the symmetry breaking scale $f$ up to several TeV. We choose
\be
f=3\tev\,,\qquad x_L = 0.5
\ee
as a benchmark value. Again the latter choice minimizes the LHT contributions to EWP observables. 

The direct bounds on mirror quarks will push their masses in the multi-TeV regime, and we choose
\be
4 \tev < m^q_H < 8 \tev\,,
\ee
with the upper bound obtained from an expected improvement on the four fermion operator constraints. 

{Before proceeding to the numerical analysis, we note that the LHT model suffers from severe fine-tuning in this case. This questions the original motivation for Little Higgs models as a natural solution to the little hierarchy problem.
However we still think that a high scale scenario is worth being considered in terms of its flavour phenomenology. In the absence of a new physics discovery at the LHC, most new physics scenarios will have a severe fine-tuning problem and the naturalness hypothesis will be challenged. In this case it will be important to question the concept of naturalness as one of our main guiding principles. No stone should be left unturned in the search for new physics, even if a model seems theoretically less motivated. In this spirit we consider it worth investigating whether in the absence of a NP signal in direct searches and Higgs data, flavour violating decays can still show a significant deviation from the SM prediction.}

\section{Numerical analysis } \label{sec:num}
\subsection{Strategy}

\begin{table}[!tb]
\center{\begin{tabular}{|l|l|}
\hline
$G_F = 1.16638(1)\cdot 10^{-5}\gev^{-2}$\hfill\cite{Agashe:2014kda} 	&  $m_{B_d}=5279.58(17)\mev$\hfill\cite{Agashe:2014kda}\\
$M_W = 80.385(15) \gev$\hfill\cite{Agashe:2014kda}  								&	$m_{B_s} =
5366.8(2)\mev$\hfill\cite{Agashe:2014kda}\\
$\sin^2\theta_W = 0.23126(13)$\hfill\cite{Agashe:2014kda} 				& 	$F_{B_d} =
190.5(42)\mev$\hfill \cite{Aoki:2013ldr}\\
$\alpha(M_Z) = 1/127.9$\hfill\cite{Agashe:2014kda}									& 	$F_{B_s} =
227.7(45)\mev$\hfill \cite{Aoki:2013ldr}\\
$\alpha_s(M_Z)= 0.1185(6) $\hfill\cite{Agashe:2014kda}			&  $\hat B_{B_d} =1.27(10)$,  $\hat
B_{B_s} =
1.33(6)$\;\hfill\cite{Aoki:2013ldr}
\\\cline{1-1}
$m_u(2\gev)=2.16(11)\mev $ 	\hfill\cite{Aoki:2013ldr}						& 
$\hat B_{B_s}/\hat B_{B_d}
= 1.06(11)$ \hfill \hfill\cite{Aoki:2013ldr} \\
$m_d(2\gev)=4.68(16)\mev$	\hfill\cite{Aoki:2013ldr}							& 
$F_{B_d} \sqrt{\hat 
B_{B_d}} = 216(15)\mev$\hfill\cite{Aoki:2013ldr} \\
$m_s(2\gev)=93.8(24) \mev$	\hfill\cite{Aoki:2013ldr}				&
$F_{B_s} \sqrt{\hat B_{B_s}} =
266(18)\mev$\hfill\cite{Aoki:2013ldr} \\
$m_c(m_c) = 1.275(25) \gev$ \hfill\cite{Agashe:2014kda}					&
$\xi =
1.268(63)$\hfill\cite{Aoki:2013ldr} \\
$m_b(m_b)=4.18(3)\gev$\hfill\cite{Agashe:2014kda} 			& $\eta_B=0.55(1)$\hfill\cite{Buras:1990fn,Urban:1997gw}
\\
$m_t(m_t) = 163(1)\gev$\hfill\cite{Laiho:2009eu,Allison:2008xk} 
						&  $\Delta M_d = 0.510(3)
\,\text{ps}^{-1}$\hfill\cite{Amhis:2014hma}\\\cline{1-1}
$m_K= 497.614(24)\mev$	\hfill\cite{Agashe:2014kda}								&  $\Delta M_s = 17.757(21)
\,\text{ps}^{-1}$\hfill\cite{Amhis:2014hma}
\\	
$F_K = 156.1(11)\mev$\hfill\cite{Laiho:2009eu}												&
$S_{\psi K_S}= 0.691(17)$\hfill\cite{Amhis:2014hma}\\
$\hat B_K= 0.750(15)$\hfill \cite{Aoki:2013ldr,Buras:2014maa}										&
$S_{\psi\phi}= 0.015(35)$\hfill \cite{Amhis:2014hma}\\
$\kappa_\epsilon=0.94(2)$\hfill\cite{Buras:2008nn,Buras:2010pza}				& $\Delta\Gamma_s/\Gamma_s=0.122(9)$\hfill\cite{Amhis:2014hma}
\\	
$\eta_{cc}=1.87(76)$\hfill\cite{Brod:2011ty}												
	& $\tau_{B_s}= 1.509(4)\,\text{ps}$\hfill\cite{Amhis:2014hma}\\		
$\eta_{tt}=0.5765(65)$\hfill\cite{Buras:1990fn}												
& $\tau_{B_d}= 1.520(4) \,\text{ps}$\hfill\cite{Amhis:2014hma}\\
$\eta_{ct}= 0.496(47)$\hfill\cite{Brod:2010mj}												
& $\tau_{B^\pm}= 1.638(4)\,\text{ps}$\hfill\cite{Amhis:2014hma}    \\
$\Delta M_K= 0.5293(9)\cdot 10^{-2} \,\text{ps}^{-1}$\hfill\cite{Agashe:2014kda}	&
$|V_{us}|=0.2253(8)$\hfill\cite{Agashe:2014kda}\\
$|\eps_K|= 2.228(11)\cdot 10^{-3}$\hfill\cite{Agashe:2014kda}					& $\gamma =(73.2^{+6.3}_{-7.0})^\circ$
\hfill\cite{Trabelsi:2014}\\
  $|V^\text{\rm avg}_{cb}|=40.7(14)\cdot 10^{-3}$\hfill \cite{Buras:2015qea} &
$|V^\text{\rm avg}_{ub}|=3.88(29)\cdot 10^{-3}$\hfill \cite{Buras:2015qea}\\
\hline
\end{tabular}  }
\caption {\textit{Values of the experimental and theoretical
    quantities used as input parameters  as of July 2015. For future 
updates see PDG \cite{Agashe:2014kda}, FLAG \cite{Aoki:2013ldr} and HFAG \cite{Amhis:2014hma}. }}
\label{tab:input}
\end{table}

An important part of our analysis is the choice of the values of CKM parameters 
as this specifies the room left for NP contributions. We will use 
the CKM parameters determined in tree level decays. These are 
\be\label{trees}
\vus, \qquad \vub, \qquad \vcb, \qquad \gamma \,.
\ee
The values for $\vus$ and the angle $\gamma$ used by us are 
\cite{Amhis:2014hma,Trabelsi:2014}
\begin{equation}\label{gamma}
  |V_{us}| = 0.2253\pm 0.0008,\qquad   \gamma = (73.2^{+6.3}_{-7.0})^\circ.
\end{equation}

The status of $\vcb$ is  not satisfactory, with exclusive 
determinations \cite{Bailey:2014tva,Bailey:2014bea,Aoki:2013ldr} giving significantly lower values than the inclusive \cite{Alberti:2014yda} ones
\be\label{vcb}
\vcb_\text{excl}=(39.36\pm0.75)\cdot 10^{-3},\qquad \vcb_\text{incl}=(42.21\pm0.78)\cdot 10^{-3}\,,
\ee
implying  the  weighted average of these results provided in \cite{Buras:2015qea} 
\be\label{vcbaverage}
\vcb_{\rm avg}=(40.7\pm 1.4)\cdot 10^{-3}
\ee
that we will adopt in what follows.

The status of $\vub$ is even worse  due to the
tensions between exclusive \cite{Bailey:2014bea} and inclusive \cite{Aoki:2013ldr} determinations of $\vub$:
\be
\vub_\text{excl}=(3.72\pm0.14)\cdot 10^{-3},\qquad 
\vub_\text{incl}=(4.40\pm0.25)\cdot 10^{-3}.
\ee

The  weighted average of these results provided in \cite{Buras:2015qea} reads
\be\label{average}
\vub_{\rm avg} =(3.88\pm0.29)\cdot 10^{-3}, 
\ee 
but due to the recent LHCb result which gives the even lower value of  $\vub=3.25\cdot 10^{-3}$ the situation is rather unclear. For the time being we will use the value in (\ref{average}). 

{In this context, it should be mentioned that, using the central values of other 
input parameters, even with the inclusive value 
of $\vcb$, the value of $\varepsilon_K$ in the SM is typically by $(10-20)\%$ 
below the data, unless the high inclusive value of $\vub$ is used. However, the 
large uncertainty in $\eta_{cc}$ found at NNLO level in \cite{Brod:2011ty} implies an uncertainty of roughly $\pm 6\%$ in $\varepsilon_K$ softening the tension in the SM with $\varepsilon_K$. For a recent discussion see \cite{Buras:2013raa}.

On the other hand for $\vub\ge 3.6\cdot 10^{-3}$  the asymmetry $S_{\psi K_S}$ 
predicted by the SM  is larger than its experimental value. For the 
inclusive value of $\vub$ it is even by $3\sigma$  above the 
data. Then  new CP phases in the $B_d^0-\bar B_d^0$ system are required to achieve an agreement with experiment, while then $\varepsilon_K$ in the SM is 
fully consistent with the data. Thus some tension between the values of $\varepsilon_K$ and  $S_{\psi K_S}$ in the SM is still present \cite{Lunghi:2008aa,Buras:2008nn}, but to reach a final conclusion, much higher accuracies on $\vub$, $\vcb$ 
and also on $\eta_{cc}$ are required.}

The remaining input parameters are collected in Table~\ref{tab:input}. We will 
comment on some of them whenever necessary.
{For the new parameters of the LHT model we will impose the bounds summarised 
in Section~\ref{sec:bounds}. As in our 2009 analysis \cite{Blanke:2009am} we perform a randomised numerical scan over the LHT parameter space, varying the input parameters in their $1\sigma$ ranges. For both scenarios A and B we generate a set of 10,000 parameter points each that satisfy the present $\Delta F = 2$ constraints at the $1\sigma$ level.}

\subsection{Results for Scenario A}

\boldmath
\subsubsection{$\Delta F=2$ constraints} 
\unboldmath

{The presence of new contributions to the $\Delta F =2$ observables in  (\ref{DF2}) and (\ref{SDATA}) allows to resolve possible tensions present in the SM, thereby putting 
some constraints on the new parameters. These $\Delta F=2$ constraints will be taken into account in the predictions for $\Delta F=1$ observables presented 
below. 

\begin{figure}
\centering
\includegraphics[width=.55\textwidth]{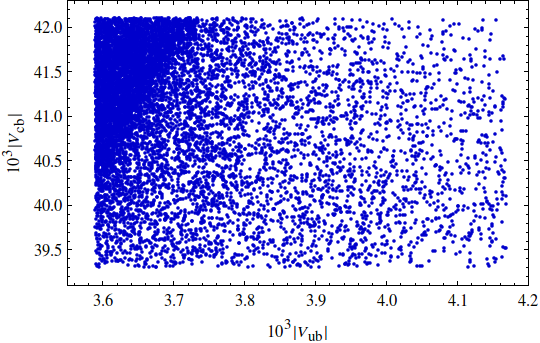}
\caption{\label{fig:VubVcb}\it Distribution of viable LHT parameter points ($f=1\tev$) in the $|V_{ub}|,|V_{cb}|$ plane, obtained with a prior flat distribution.}
\end{figure}

At this point it should be  recalled that in the LHT model {the CP asymmetry $S_{\psi\phi}$ can both be enhanced and suppressed w.\,r.\,t.\ the SM.}
 We will see this in the figures below. This is not always the case in other models. For 
instance in the Two Higgs Doublet Model with MFV {and flavour blind phases} (${\rm 2HDM_{\overline{MFV}}}$)  \cite{Buras:2010mh,Buras:2010zm}, the asymmetry  $S_{\psi\phi}$ can only be {enhanced} due to its correlation with $S_{\psi K_S}$. { Thus if eventually $S_{\psi\phi} < (S_{\psi\phi})_\text{SM}$ will be found, the LHT model will still be viable, in contrast to the ${\rm 2HDM_{\overline{MFV}}}$.}

Fig.~\ref{fig:VubVcb} demonstrates that the LHT model can fit the data on $\Delta F=2$ observables for the full range of the measured values of $\vub$ and $\vcb$ covered in our scan.
Yet, small values of $\vub$ and large values of $\vcb$ are favoured as 
for such values the data on $S_{\psi K_S}$ and $\varepsilon_K$ are easiest 
to satisfy, respectively.

\boldmath
\subsubsection{$\kpn$ and $\klpn$}
\unboldmath

\begin{figure}
\centering
\includegraphics[width=.55\textwidth]{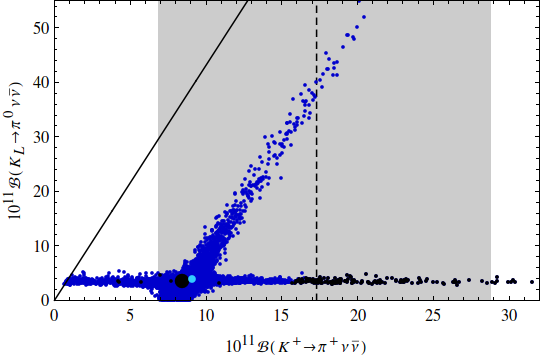}
\caption{\label{fig:Kpinunu}\it Correlation between the branching ratios of $K^+\to\pi^+\nu\bar\nu$ and $K_L\to\pi^0\nu\bar\nu$ in the LHT model for $f=1\tev$. The large black dot shows the central SM value for our choice of input parameters, and the light blue point shows the contribution from the T-even sector. The black LHT points are excluded by the constraint from $K_L\to\mu^+\mu^-$ \cite{Isidori:2003ts}. The experimental $1\sigma$ range for $\mathcal{B}(K^+\to\pi^+\nu\bar\nu)$ \cite{Artamonov:2008qb} is displayed by the grey band, while the solid black line indicates the Grossman-Nir bound \cite{Grossman:1997sk}.}
\end{figure}

The correlation between $\kpn$ and  $\klpn$ has been the subject of many analyses. In Fig.~\ref{fig:Kpinunu} we show the correlation between $\mathcal{B}(\kpn)$ and 
$\mathcal{B}(\klpn)$ as obtained from the randomised scan over the LHT parameters. The experimental
$1\sigma$-range for $\mathcal{B}(\kpn)$ \cite{Artamonov:2008qb} and the
model-independent Grossman-Nir (GN) bound \cite{Grossman:1997sk} are also
shown. We observe that the two
branches of possible points found in \cite{Blanke:2006eb} are still present and that significant enhancements with respect to the SM predictions {are allowed. In fact the possible enhancements are larger than in our 2009 analysis. This counter-intuitive result originates in the non-decoupling behaviour of the mirror quarks which, due to the constraints from LHC Run 1, have to be heavier than assumed by us six years ago.} The first branch, which is parallel to the GN-bound, leads to possible large enhancements in $\mathcal{B}(\klpn)$ so that,  without the constraint from $\epe$, values
as high as $5\cdot 10^{-10}$ are possible, being at the same time
consistent with the measured value for $\mathcal{B}(\kpn)$. The latter branching
 ratio can reach values in the ballpark of $2\cdot 10^{-10}$. 
On the second branch, which corresponds
to values for $\mathcal{B}(\klpn)$ rather close to its SM prediction,
 $\mathcal{B}(\kpn)$ can be strongly suppressed but also enhanced. However the 
size of this enhancement is limited by the $K_L\to \mu^+\mu^-$ constraint so 
that  the present central experimental value can only barely be   reached. 
We will return to this constraint in explicit terms below.

The presence of the two branches is a remnant of the specific operator structure of the LHT model and has been analysed in a model-independent manner in \cite{Blanke:2009pq}. Consequently observing one day the $K\to\pi\nu\bar\nu$ branching ratios outside these two branches would not only rule out the LHT model but at the same time put all models with a similar flavour structure in difficulties. On the other hand in models like the custodially protected Randall-Sundrum (RS) model in which new flavour violating operators are present, no visible correlation is observed, so that an observation of the $K\to\pi\nu\bar\nu$ modes outside the two branches can be explained in such kind of models \cite{Blanke:2008yr}. This is also possible in models with tree-level flavour-violating $Z$ and $Z^\prime$ exchanges \cite{Buras:2012jb,Buras:2015yca} if flavour changing left- and right-handed couplings are present.

\boldmath
\subsubsection{$\klpn$,  $S_{\psi K_S}$ and  $S_{\psi\phi}$.}
\unboldmath

\begin{figure}
\centering
\includegraphics[width=.55\textwidth]{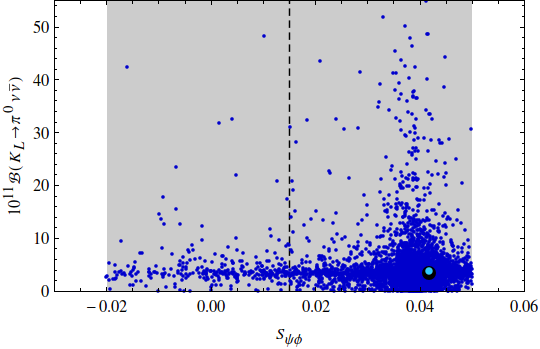}
\caption{\label{fig:Spsiphi-KL}\it Branching ratio of $K_L\to\pi^0\nu\bar\nu$ as a function of $S_{\psi \phi}$ in the LHT model for $f=1\tev$. The large black dot shows the central SM value for our choice of input parameters, and the light blue point shows the contribution from the T-even sector. The experimental $1\sigma$ range for $S_{\psi \phi}$ is displayed by the grey band \cite{Amhis:2014hma}.
}
\end{figure}

Next, of particular interest are the correlations of $\klpn$ with
the asymmetries   $S_{\psi K_S}$ and $S_{\psi\phi}$. In 2009 we have pointed out that large departures of  $S_{\psi\phi}$
 from its SM value would not allow for large NP effects in the rare $K$ decay within the LHT model. But as seen in (\ref{SDATA}) the present 
experimental value for this asymmetry fully agrees with the SM.
In Fig.~\ref{fig:Spsiphi-KL} we show the correlation of  $\mathcal{B}(\klpn)$
 with  $S_{\psi\phi}$. We observe that within the LHT model  $S_{\psi\phi}$ 
can still differ significantly from its SM value of $0.04$ but 
large enhancements of  $\mathcal{B}(\klpn)$  are {most likely  when 
 $S_{\psi\phi}$  is SM-like. It should also be noted that the large new physics 
effects are due to mirror fermions as the T-even sector is CMFV like.}

 \begin{figure}
\centering
\includegraphics[width=.55\textwidth]{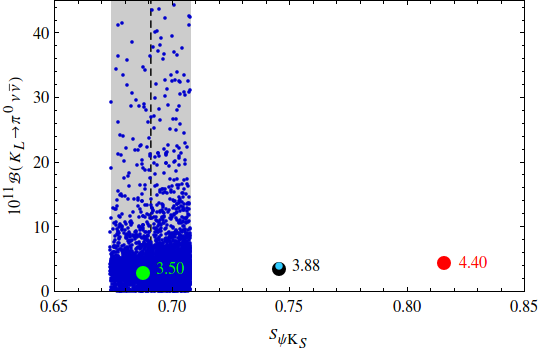}
\caption{\label{fig:SpsiKS-KL}\it Branching ratio of $K_L\to\pi^0\nu\bar\nu$ as a function of $S_{\psi K_S}$ in the LHT model for $f=1\tev$. The large black dot shows the central SM value for our choice of input parameters, and the light blue point shows the contribution from the T-even sector. The green and red dots indicate the SM predictions for $|V_{ub}|_\text{excl.}=3.5\cdot 10^{-3}$ and $|V_{ub}|_\text{incl.}=4.4\cdot 10^{-3}$, respectively. The experimental $1\sigma$ range for $S_{\psi K_S}$ is displayed by the grey band \cite{Amhis:2014hma}.}
\end{figure}

Fig.~\ref{fig:SpsiKS-KL} demonstrates that for high values of $\vub$ the 
T-even sector would not be capable to lower the value of $S_{\psi K_S}$ to 
agree with the data, while this can be achieved with the help of  the mirror fermions simultaneously allowing for significant departures of the branching ratio  for $\klpn$ from its SM value.

\boldmath
\subsubsection{Correlation of $K\to\pi\nu\bar\nu$ with $K_L\to\mu^+\mu^-$ and 
$\epe$}
\unboldmath

Of interest are also the  correlations of  $K\to\pi\nu\bar\nu$ with 
$K_L\to\mu^+\mu^-$ and $\epe$ as they can limit possible NP effects in 
 $K\to\pi\nu\bar\nu$. 
In Fig.~\ref{fig:KLmumu} we show the correlation between  $K_L \to \mu^+ \mu^-$ and $\kpn$. As pointed out in \cite{Blanke:2008yr} this linear correlation on the upper branch should be contrasted with the inverse correlation between the two decays in question found in the custodially protected RS model. The origin of this difference is the operator structure of the models in question: While in the LHT model rare $K$ decays are mediated as in the SM by left-handed currents, in the RS model in question the flavour violating $Z$ coupling to right-handed quarks dominates.
In the LHT model consequently a large enhancement of $\mathcal{B}(\kpn)$ automatically implies a significant 
enhancement of $\mathcal{B}(K_L \to \mu^+ \mu^-)_\text{SD}$ and this is not always allowed by 
the upper bound $\mathcal{B}(K_L \to \mu^+ \mu^-)_\text{SD}<2.5 \cdot 10^{-9}$ \cite{Isidori:2003ts}, displayed by the dotted line in  Fig.~\ref{fig:KLmumu}. The horizontal branch 
in this figure, on which  $\mathcal{B}(\kpn)$ is not constrained by $K_L\to\mu^+\mu^-$, corresponds to the upper branch in Fig.~\ref{fig:Kpinunu}, while the 
upper one in   Fig.~\ref{fig:KLmumu} to the lower one in  Fig.~\ref{fig:Kpinunu}.

\begin{figure}
\centering
\includegraphics[width=.55\textwidth]{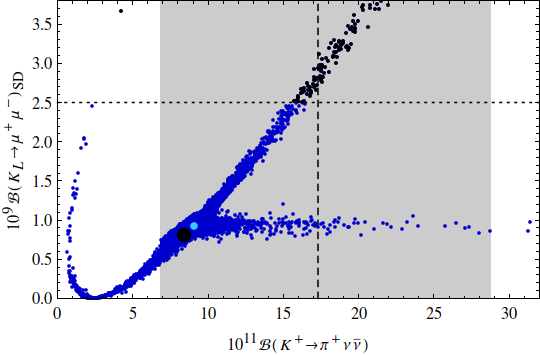}
\caption{\label{fig:KLmumu}\it Correlation between the short-distance contribution to $\mathcal{B}(K_L\to\mu^+\mu^-)$ and the branching ratio of $K^+\to\pi^+\nu\bar\nu$ in the LHT model for $f=1\tev$. The large black dot shows the central SM value for our choice of input parameters, and the light blue point shows the contribution from the T-even sector. The black LHT points are excluded by the constraint from $K_L\to\mu^+\mu^-$, indicated by the horizontal dotted line \cite{Isidori:2003ts}. The experimental $1\sigma$ range for $\mathcal{B}(K^+\to\pi^+\nu\bar\nu)$ is displayed by the grey band \cite{Artamonov:2008qb}.}
\end{figure}

Another interesting correlation is the one of $\klpn$ and $\epe$ which has 
been analysed by us in the LHT model in \cite{Blanke:2007wr}. As we 
summarised in Section~\ref{sec:epe} significant progress has been made since 
then both by lattice QCD and large $N$ through the improved determination of 
the relevant hadronic matrix elements of QCD and electroweak penguin 
operators. Using the upper bound on $\bsi$ and $\bei$ in (\ref{BG15}) the authors of {\cite{Buras:2015yba} find
$\epe$ in the SM at the bound in (\ref{SMbound}) to be roughly by $ 2\sigma$  lower than the data.}

\begin{figure}
\centering
\includegraphics[width=.55\textwidth]{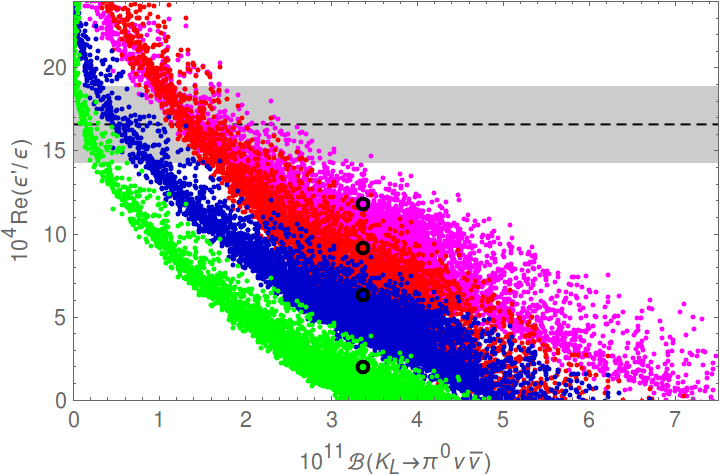}
\caption{\label{fig:epe}\it Correlation between $\mathcal{B}(K_L\to\pi^0\nu\bar\nu)$ and $\RE(\epe)$ in the LHT model for $f=1\tev$ for different values of $(\bsi,\bei)$: $(1.0,1.0)$ (red), $(0.76, 0.76)$ (blue), $(0.57,0.76)$ (green), $(1.0,0.76)$ (magenta). The black dots show the corresponding central SM values. The experimental $1\sigma$ range for $\RE(\epe)$ is displayed by the grey band \cite{Agashe:2014kda,Batley:2002gn,AlaviHarati:2002ye,Worcester:2009qt}.}
\end{figure}

In our analysis we will consider {first of all} three choices for the pair $(\bsi,\bei)$:
\be\label{BGbound}
\bsi=\bei =1.0, \qquad (\text{red}),
\ee
corresponding to the upper bound in  (\ref{BG15}),
\be
\bsi=\bei= 0.76,  \qquad (\text{blue}),
\ee
corresponding to the central lattice value for $\bei$ and the largest 
value for $\bsi$ consistent with the bound in  (\ref{BG15})
and 
\be
\bsi=0.57, \qquad \bei=0.76  \qquad (\text{green})
\ee
corresponding to the central lattice values.

{In Fig.~\ref{fig:epe} we show the correlation between $\klpn$ and $\epe$ for these three scenarios. We observe that in the second and third  case the SM prediction is significantly below the data. Requiring the LHT model to obtain agreement with the data suppresses strongly the branching ratio $\mathcal{B}(K_L\to\pi^0\nu\bar\nu)$ below its SM value. At the bound in (\ref{BGbound})  taking all the uncertainties into account the suppression is moderate. This is in particular the case
if we allow to violate the inequality between $\bsi$ and $\bei$ and choose 
\be
\bsi=1.0, \qquad \bei=0.76  \qquad (\text{magenta})\,.
\ee
But this case is very unlikely in view of the bound in (\ref{BG15}).

\begin{figure}
\centering
\includegraphics[width=.55\textwidth]{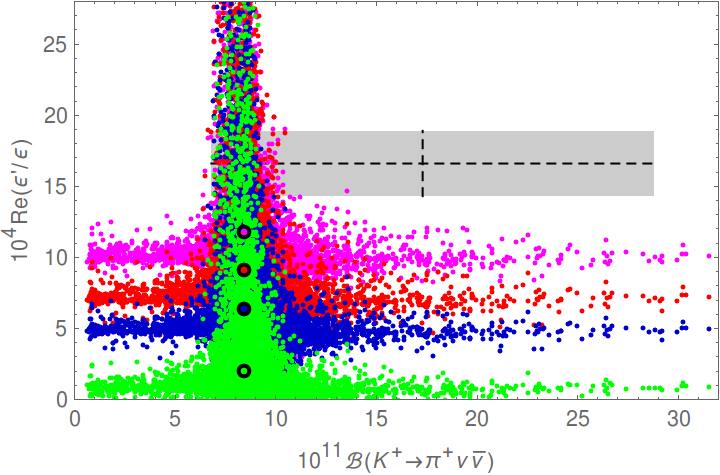}
\caption{\label{fig:epe+}\it Correlation between $\mathcal{B}(K^+\to\pi^+\nu\bar\nu)$ and $\RE(\epe)$ in the LHT model for $f=1\tev$ for different values of $(\bsi,\bei)$: $(1.0,1.0)$ (red), $(0.76, 0.76)$ (blue), $(0.57,0.76)$ (green), $(1.0,0.76)$ (magenta). The black dots show the corresponding central SM values. The experimental $1\sigma$ ranges are displayed by the grey band \cite{Artamonov:2008qb,Agashe:2014kda,Batley:2002gn,AlaviHarati:2002ye,Worcester:2009qt}.}
\end{figure}

Fig.~\ref{fig:epe+} shows the analogous correlation between
$\mathcal{B}(K^+\to\pi^+\nu\bar\nu)$ and $\RE(\epe)$. The two branches of Fig.~\ref{fig:Kpinunu} also manifest themselves in the present figure. The horizontal branch with large enhancements of $\mathcal{B}(K^+\to\pi^+\nu\bar\nu)$ is disfavoured by $\epe$. Fitting the data on $\epe$ is possible within the LHT model without any suppression of $\mathcal{B}(K^+\to\pi^+\nu\bar\nu)$. However significant 
modifications of this branching ratio with respect to the SM are then not allowed.}

\boldmath
\subsubsection{Problems with $B_{s,d}\to\mu^+\mu^-$ and $B_d\to K^{(*)} \ell^+\ell^-$}
\unboldmath

While until now the LHT model passed all experimental tests related to $\Delta F=2$ 
transitions and rare $K$ decays, the situation changes when 
$B_{s,d}\to\mu^+\mu^-$ and $B_d\to K^{(*)} \ell^+\ell^-$ are considered. 

\begin{figure}
\centering
\includegraphics[width=.55\textwidth]{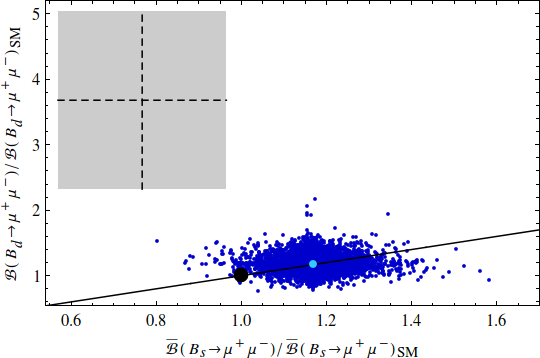}
\caption{\label{fig:Bsdmumu-1TeV}\it Correlation between $\bar{\mathcal{B}}(B_s\to\mu^+\mu^-)$ and $\mathcal{B}(B_d\to\mu^+\mu^-)$ in the LHT model for $f=1\tev$. The large black dot shows the central SM value for our choice of input parameters, and the light blue point shows the contribution from the T-even sector. The experimental $1\sigma$ ranges are displayed by the grey rectangle \cite{CMS:2014xfa}, and the MFV prediction is indicated by the solid black line.}
\end{figure}

In Fig.~\ref{fig:Bsdmumu-1TeV}  we show the correlation between the ratios $\mathcal R_{s,d}^{\mu\mu}$ in the LHT model. While the MFV prediction, represented
 by the straight black line, can be modified, this modification {is not sufficient to bring the theory in full agreement with the data. While the data would favour a suppression of $\mathcal{B}(B_s\to\mu^+\mu^-)$  relative to its SM value, the LHT model favours its enhancement. The contribution from the T-even sector provides a flavour universal enhancement by $15\%$, and particular values of model parameters in the T-odd sector are required to change this pattern. We find that while the mirror quarks can enhance $\mathcal{B}(B_d\to\mu^+\mu^-)$ by up to a factor of 2, such large values appear impossible together with a suppression of $\mathcal{B}(B_s\to\mu^+\mu^-)$. Consequently finding future data to 
confirm the present ranges of $\mathcal{B}(B_{s,d}\to\mu^+\mu^-)$ will be problematic for the LHT model.

\begin{figure}
\centering
\includegraphics[width=.55\textwidth]{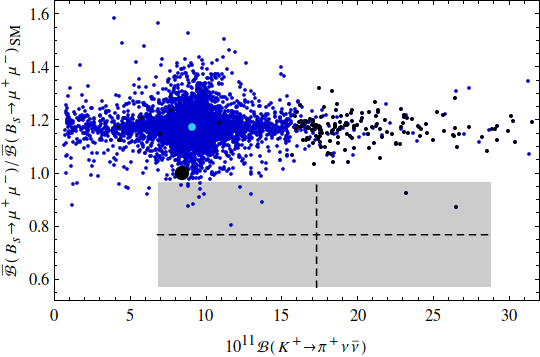}
\caption{\label{fig:KPBsmumu}\it Correlation between $\mathcal{B}(K^+\to\pi^+\nu\bar\nu)$ and $\bar{\mathcal{B}}(B_s\to\mu^+\mu^-)$   in the LHT model for $f=1\tev$. The large black dot shows the central SM value for our choice of input parameters, and the light blue point shows the contribution from the T-even sector. The experimental $1\sigma$ ranges are displayed by the grey rectangle \cite{CMS:2014xfa,Artamonov:2008qb}. The black LHT points are excluded by the constraint from $K_L\to\mu^+\mu^-$.}
\end{figure}

The difficulty to suppress $\mathcal{B}(B_s\to\mu^+\mu^-)$ below its SM value is also seen in Fig.~\ref{fig:KPBsmumu}. Additionally we observe that for the lowest values of $\mathcal{B}(B_s\to\mu^+\mu^-)$ favoured by the data, 
large enhancements of $\mathcal{B}(\kpn)$ are not allowed.}

{Even more problematic for the LHT model appear at present the data on $B_d\to K(K^*) \ell^+\ell^-$ as we discussed already in section \ref{sec:bsll}.}

\boldmath
\subsection{\boldmath${B\to K^{(*)}\nu\bar\nu}$\unboldmath}
\unboldmath

\begin{figure}
\centering
\includegraphics[width=.55\textwidth]{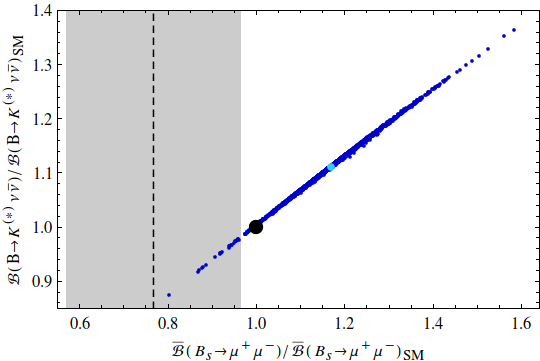}
\caption{\label{fig:Bnunu}\it Correlation between $\bar{\mathcal{B}}(B_s\to\mu^+\mu^-)$ and $\mathcal{B}(B\to K^{(*)}\nu\bar\nu)$ in the LHT model for $f=1\tev$. The large black dot shows the central SM value for our choice of input parameters, and the light blue point shows the contribution from the T-even sector. The experimental $1\sigma$ range for $\bar{\mathcal{B}}(B_s\to\mu^+\mu^-)$ is displayed by the grey band \cite{CMS:2014xfa}.}
\end{figure}

In Fig.~\ref{fig:Bnunu} we show the correlation between 
$\bar{\mathcal{B}}(B_s\to\mu^+\mu^-)$ and $\mathcal{B}(B\to K^{(*)}\nu\bar\nu)$ in the LHT model. We observe a very strong {linear} correlation characteristic 
for models with {left-handed flavour changing currents in which the $Z$ penguin dominates}. We also note as in Fig.~\ref{fig:KPBsmumu} that the T-even sector by itself would be in conflict with experiment 
but the presence of mirror quarks allows still to save the LHT model.
{Yet, as already seen in Fig.~\ref{fig:KPBsmumu}, it is difficult to obtain results within $1\sigma$ from the experimental central value.}

\boldmath
\subsection{Results for Scenario B }
\unboldmath
 
{Let us finally study the pessimistic scenario that no new particles will be discovered at the LHC and all electroweak and Higgs physics observables turn out to be SM-like. In this case the symmetry breaking scale $f$ and the mirror fermion masses will be pushed into the multi-TeV range, as discussed in section \ref{sec:sceB}. 

\begin{figure}
\centering
\includegraphics[width=.55\textwidth]{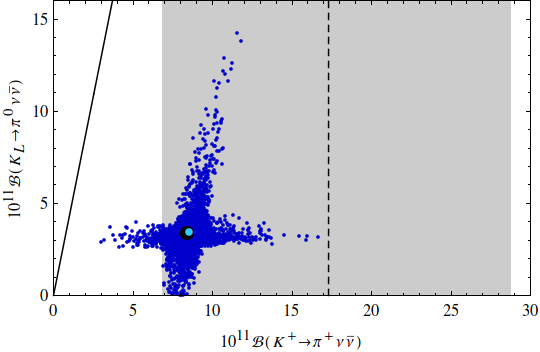}
\caption{\label{fig:Kpinunu-3TeV}\it Correlation between the branching ratios of $K^+\to\pi^+\nu\bar\nu$ and $K_L\to\pi^0\nu\bar\nu$ in the LHT model for $f=3\tev$. The large black dot shows the central SM value for our choice of input parameters, and the light blue point shows the contribution from the T-even sector. The experimental $1\sigma$ range for $\mathcal{B}(K^+\to\pi^+\nu\bar\nu)$ is displayed by the grey band \cite{Artamonov:2008qb}, while the solid black line indicates the Grossman-Nir bound \cite{Grossman:1997sk}.}
\end{figure}

It turns out that in this case rare $K$ decays, in particular the $K \to \pi\nu\bar\nu$ decays, are the best channels to observe a sign of the LHT model. As we can see in Fig.\ \ref{fig:Kpinunu-3TeV},
significant enhancements of the branching ratios of $K^+\to\pi^+\nu\bar\nu$ and $K_L\to\pi^0 \nu\bar\nu$  will still be possible. Again we observe the known two-branch structure. On the horizontal branch $K_L\to
\pi^0 \nu\bar\nu$ remains SM-like, while $K^+\to\pi^+\nu\bar\nu$ can be enhanced by up to a factor of two. On the second branch the impact on $K^+\to\pi^+\nu\bar\nu$ is more modest, but $\mathcal{B}(K_L\to\pi^0 \nu\bar\nu)$ can be larger than its SM prediction by up to a factor of four.
But again if the present low values of $\RE(\epe)_\text{SM}$ will be confirmed by more precise lattice calculations, only a suppression of $\mathcal{B}(\klpn)$ in Fig.\ \ref{fig:Kpinunu-3TeV} will be allowed and $\mathcal{B}(\kpn)$ will be SM-like.

\begin{figure}
\centering
\includegraphics[width=.55\textwidth]{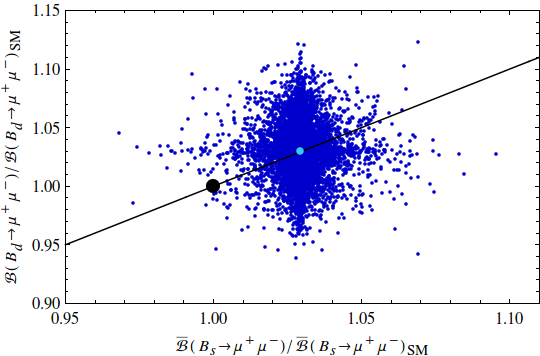}
\caption{\label{fig:Bsdmumu-3TeV}\it Correlation between $\bar{\mathcal{B}}(B_s\to\mu^+\mu^-)$ and $\mathcal{B}(B_d\to\mu^+\mu^-)$ in the LHT model for $f=3\tev$. The large black dot shows the central SM value for our choice of input parameters, and the light blue point shows the contribution from the T-even sector. }
\end{figure}

The effects in rare $B$ decays on the other hand turn out to be much smaller and, in view of experimental and parametric uncertainties, will be difficult to disentangle from the SM. In Fig.\ \ref{fig:Bsdmumu-3TeV} we show the correlation between $\bar{\mathcal{B}}(B_s\to\mu^+\mu^-)$ and $\mathcal{B}(B_d\to\mu^+\mu^-)$ as an example.

It is interesting to see how the LHT effects in rare meson decays scale with the symmetry breaking scale $f$. Naively, the new contributions are suppressed by $v^2/f^2$ with respect to the SM. This is indeed what we see in the T-even sector, displayed by the light blue point in the figures. The case of the T-odd sector is however a bit more involved. Firstly, as we increase the mirror quark masses simultaneously with the scale $f$, the size of the loop functions remains unchanged and the only suppression stems from the $v^2/f^2$ prefactor. However simultaneously the constraints on the T-odd sector from $\Delta F = 2$ observables become weaker, they scale like
\be
\xi^2 \frac{v^2}{f^2} < \epsilon
\ee
with $\xi$ denoting the relevant combination of $V_{Hd}$ elements and $\epsilon \ll 1$ depending on the meson sector in question. The T-odd contributions to $\Delta F = 1$ processes on the other hand scale as
\be
\xi \frac{v^2}{f^2} < \frac{v}{f} \sqrt{\epsilon}\,.
\ee
We conclude that the mirror quark contributions are only linearly suppressed by the scale $f$. 

}

\section{Summary}\label{sec:sum}

In this paper we have presented a new analysis of quark
flavour observables within the LHT model. Our analysis takes into account the  
most recent data from the LHCb experiment, the improvements on CKM parameters and hadronic parameters from lattice QCD and the new lower bounds on the masses 
of new gauge bosons and mirror quarks. Our main findings are as follows:
\begin{itemize}
\item
The LHT model agrees well with the data on $\Delta F=2$ observables and is 
capable of removing some slight tensions between the SM predictions and the data.
\item
The most interesting departures from SM predictions can be found for $\kpn$ and 
$\klpn$ decays, when only constraints from  $\Delta F=2$ observables are taken 
into account. {An enhancement} of the branching ratio for $\kpn$  by a factor of two relative to the SM prediction \cite{Buras:2015qea} is {still} possible. An even 
larger enhancement in the case of $\klpn$ is allowed. {But as we have 
shown in Fig.~\ref{fig:epe}, the recent analysis of $\epe$ in the SM {\cite{Buras:2015yba}}, based 
on new results for the non-perturbative parameters $\bsi$ and $\bei$ from 
lattice QCD \cite{Blum:2015ywa,Bai:2015nea} and the large $N$ approach {\cite{Buras:2015xba}},
appears to exclude this possibility at present. Rather a suppression of  $\klpn$ is required to fit the data on $\epe$. On the other hand 
as seen in  Fig.~\ref{fig:epe+}, no significant shifts of $\kpn$ with respect to SM are allowed. }
\item
NP effects in rare $B_{s,d}$ decays are significantly smaller than in rare $K$ 
decays. { Still they can amount to up to a factor of 2 in the $b\to d$ system and to about $50\%$ of the SM branching ratios in $b\to s$ transitions, like $\mathcal{B}(B_s\to\mu^+\mu^-)$ and $B\to K^{(*)}\nu\bar\nu$. } 
\item
{More interestingly} the pattern of departures from SM expectations for $B_{s,d}$ decays predicted by the LHT model 
disagrees with the present data. $\mathcal{B}(B_s\to\mu^+\mu^-)$ is favoured 
by this model to be enhanced rather than suppressed as indicated by the data, and the simultaneous enhancement of  $\mathcal{B}(B_d\to\mu^+\mu^-)$ cannot be explained.
{Furthermore, the LHT model fails to reproduce the 
$B_d\to K^{(*)}\ell^+\ell^-$  and $R(D^{(*)})$ anomalies observed by the LHCb, BaBar and Belle experiments.}
\end{itemize}

The future of the LHT model depends crucially on the improved experimental 
values of  $\mathcal{B}(B_{s,d}\to\mu^+\mu^-)$ and on the future of the 
$B_d\to K^{(*)} \ell^+\ell^-$ anomalies. If these anomalies will be confirmed 
by future more accurate data and theory predictions, then the LHT model is not the 
NP realised by nature. For this model to survive the flavour tests in the quark sector, the 
anomalies in question have to disappear. Then also significant enhancements of the branching 
ratios for $\kpn$ and $\klpn$ will be possible. {Note however that these could be forbidden by $\epe$ if the future more precise lattice calculations of $\bsi$ confirm the bound (\ref{BG15}).}

We have also analysed the case of a higher scale $f= 3\tev$. As seen in 
Figs.~\ref{fig:Kpinunu-3TeV} and \ref{fig:Bsdmumu-3TeV}, NP effects 
are significantly smaller than for  $f= 1\tev$. {Yet the rare $K\to\pi\nu\bar\nu$ decays still show sizeable LHT effects, which are particularly welcome as in such a scenario an LHT discovery based on direct searches and electroweak and Higgs physics will be difficult. In addition thanks to the pattern of deviations a {distinction} between the SM, the LHT model, and other NP scenarios on the basis of 
flavour observables discussed by us should in principle be possible. }

In view of these definite findings
we are looking forward to improved experimental data and improved lattice 
calculations. The plots presented by us should facilitate monitoring 
the future confrontations of  the LHT model  with the data 
and help to determine whether this simple model can satisfactorily describe 
the observables considered  by us.

\section*{Acknowledgements}

This research was done in the context of the ERC Advanced Grant project ``FLAVOUR''(267104) and was partially supported by the Munich Institute for Astro- and Particle Physics (MIAPP) and the DFG cluster
of excellence ``Origin and Structure of the Universe''.

\bibliographystyle{JHEP}
\bibliography{allrefsLHT}
\end{document}